\documentclass[review,number,sort&compress]{elsarticle}
\usepackage[utf8]{inputenc} % special characters
\usepackage[english]{babel}
\usepackage{graphicx}
\usepackage{hyperref}
\usepackage{subfig}
\usepackage{xcolor}
\usepackage[colorinlistoftodos]{todonotes}

\graphicspath{{./figures/}}

\begin{document}

\begin{frontmatter}

\title{Deep Diffused APDs for Charged Particle Timing Applications: Performance after Neutron Irradiation}

\author[cern]{M.~Centis~Vignali\corref{corrauthor}}
\cortext[corrauthor]{Corresponding author}
\ead{matteo.centis.vignali@cern.ch}

\author[cern,lip]{M.~Gallinaro}
\author[uprinc]{B.~Harrop}
\author[uprinc]{C.~Lu}
\author[rmd]{M.~McClish}
\author[cern]{M.~Moll}
\author[upenn]{F.~M.~Newcomer}
\author[cern,usan]{S.~Otero~Ugobono}
\author[cern,uvirg]{S.~White}

\address[cern]{CERN, Geneva, Switzerland}
\address[lip]{LIP, Lisbon, Portugal}
\address[uprinc]{Princeton University, Princeton, USA}
\address[rmd]{Radiation Monitoring Devices, Watertown, USA}
\address[upenn]{University of Pennsylvania, Philadelphia, USA}
\address[usan]{Universidade de Santiago de Compostela, Santiago de Compostela, Spain}
\address[uvirg]{University of Virginia, Charlottesville, USA}

\begin{abstract}
%% The high luminosity upgrade of the CERN Large Hadron Collider (HL-LHC) will result in a pile-up of around 200 proton-proton collisions per bunch crossing.
%% In order to reduce the impact of pile-up on physics analyses the ATLAS and CMS experiments are planning to use the timing of the reconstructed objects within one event.
%% The primary proton-proton collisions have an RMS spread of $\approx$170~ps within one bunch crossing, and a time resolution of around 30\,ps is required for pile-up mitigation.
%% This performance requires dedicated detectors for the determination of the time of the minimum ionising particle tracks.
%% The timing detectors will be subjected, for the target integrated luminosity of 3000~fb$^{-1}$, to radiation levels corresponding to a 1-MeV neutrons fluence ($\Phi_{eq}$) of about $10^{14}$ or $10^{15}$\,cm$^{-2}$ for the barrel and end-cap detector regions of the CMS detector, respectively.

Recent interest in pile-up mitigation through fast timing at the HL-LHC has focused attention on technologies that now achieve minimum ionising particle (MIP) time resolution of 30\,picoseconds or less.
The constraints of technical maturity and radiation tolerance narrowed the options in this rapidly developing field for the ATLAS and CMS upgrades to low gain avalanche detectors and silicon photomultipliers.
In a variety of applications where occupancies and doses are lower, devices with pixel elements of order 1\,cm$^2$, nevertheless achieving 30\,ps, would be attractive.
  
In this paper, deep diffused Avalanche Photo Diodes (APDs) are examined as candidate timing detectors for HL-LHC applications.

Devices with an active area of $8 \times 8$\,mm$^2$ are characterised using a pulsed infrared laser and, in some cases, high energy particle beams.
The timing performance as well as the uniformity of response are examined.

The effects of radiation damage on current, signal amplitude, noise, and timing of the APDs are evaluated using detectors with an active area of $2 \times 2$\,mm$^2$.
These detectors were irradiated with neutrons up to a a 1-MeV neutrons fluence $\Phi_{eq} = 10^{15}$\,cm$^{-2}$.
Their timing performance was characterised using a pulsed infrared laser.

%While the large signal to noise ratio and fast rise time obtained with devices of the order $\approx 0.5$~cm$^2$ area make these devices attractive for application in regions of lower occupancy (i.e.\ central as opposed to end-cap of the LHC collider experiments),
While a time resolution of $27 \pm 1$\,ps was obtained in a beam test using an $8 \times 8$\,mm$^2$ sensor, the present study only demonstrates that gain loss can be compensated by increased detector bias up to fluences of $\Phi_{eq} = 6 \cdot 10^{13}$\,cm$^{-2}$.
So it possibly falls short of the $\Phi_{eq} = 10^{14}$\,cm$^{-2}$ requirement for the CMS barrel over the lifetime of the HL-LHC.

\end{abstract}

\end{frontmatter}

\section{Introduction}

The high luminosity upgrade of the CERN Large Hadron Collider (HL-LHC) foreseen to start in 2026 will provide an instantaneous luminosity of up to $5 \cdot 10^{34}$\,cm$^{-2}$s$^{-1}$ with a bunch spacing of 25\,ns, and an average pile-up of up to 200 collisions per bunch crossing\,\cite{hlLhcTecDesRep}.
This value of pile-up presents a challenge for the experiments, as currently ATLAS and CMS have reached a typical number of concurrent interactions within the same bunch crossing (pile-up) of approximately 35\,\cite{atlasPileup,cmsPileup}.
At present, the effects of pile-up on physics analyses are mitigated by resolving the primary vertices within one bunch crossing along the beam axis.
The physics objects are then associated to the corresponding vertices by using the tracking information.
This strategy was used also at Tevatron, where the pile-up was $\approx 6$.
The pile-up at HL-LHC will pose a challenge to this method, as a significant fraction of the primary vertices will have a distance smaller than the resolution of the vertex detectors, making an association to the reconstructed physics objects impossible.

A different method to associate the reconstructed objects to separate vertices relies on the measurement of the time of arrival of the particles at the detectors.
A sufficiently accurate time measurement effectively reduces the vertex density, improving the event reconstruction capability of the experiments.
Since the RMS spread of the primary vertices at HL-LHC is foreseen to be $\approx$ 170~ps within one bunch crossing, a minimum ionising particle (MIP) time resolution of $\approx 30$~ps is necessary to improve the association of the particles to their correct vertex, bringing the complexity of reconstruction to the present level\,\cite{cmsMIPtiming,atlasMIPtiming}.

To reach this performance, dedicated detector systems will be needed and both ATLAS and CMS are planning such upgrades\,\cite{cmsMIPtiming,atlasMIPtiming}.
The timing detectors will be subjected, for the target integrated luminosity of 3000 fb$^{-1}$, to radiation levels corresponding to a 1-MeV neutrons fluence ($\Phi_{eq}$) of about $10^{14}$ or a few $10^{15}$\,cm$^{-2}$ for the barrel and end-cap detector regions, respectively.

By exploiting the design margins of HL-LHC, it will be possible to achieve an ultimate luminosity of up to $7.5 \cdot 10^{34}$\,cm$^{-2}$s$^{-1}$ and an ultimate integrated luminosity of 4000\,fb$^{-1}$\,\cite{hlLhcTecDesRep}.
In this scenario the pile-up and fluence to which the detectors will be exposed will increase proportionally to the instantaneous and integrated luminosity, respectively.

All the proposed technologies for ATLAS and CMS timing layers involve silicon with internal gain: Silicon Photomultipliers (SiPMs) or Low Gain Avalanche Diodes (LGADs)\,\cite{cmsMIPtiming,atlasMIPtiming}.
In this context, avalanche silicon diode structures with a capacitively coupled readout for charged particle timing were evaluated.
The characteristics of these detectors make them suitable for applications where a pixel size of approximately 1\,cm$^{2}$ is appropriate, e.g.\ in the CMS barrel timing layer\,\cite{cmsMIPtiming}.
This paper summarises the characterisation of deep diffused Avalanche Photo Diodes (APDs) produced by Radiation Monitoring Devices\,\cite{rmdAddress} used as timing detectors for charged particles.
To improve the detector's timing performance, the APDs are used to directly detect the traversing particles, without a radiator medium where light is produced.
Studies of these sensors as Minimum Ionising Particle (MIP) timing detectors, using an AC-coupled readout, were performed previously and showed promising results\,\cite{white2014}.
The timing performance as well as the radiation hardness of these devices are addressed.
Section\,\ref{sec:ddApds} provides a general description of the deep diffused APDs.
In Section\,\ref{sec:samples} a detailed description of the devices used in this study is given.
Section\,\ref{sec:gain8x8led} contains the measurement of the gain of the APDs with an active area of $8 \times 8$~mm$^2$ using blue light. Both the bias and temperature dependence of the gain are explored.
Section\,\ref{sec:irrad2x2} reports the characterisation of neutron-irradiated APDs with an active area of $2 \times 2$~mm$^2$.
Sections\,\ref{sec:unif8x8laser} and\,\ref{sec:timing8x8laser}, respectively, summarise uniformity and timing measurements of APDs with an active area of $8 \times 8$~mm$^2$ with DC-coupled and AC-coupled readout.
Section\,\ref{sec:tb8x8} describes the methods and results obtained through beam tests of APDs with an active area of $8 \times 8$~mm$^2$ with AC-coupled readout.
Finally, Section\,\ref{sec:summary} summarises the results obtained in this study.

\section{Deep Diffused APDs}
\label{sec:ddApds}

Deep diffused APDs consist of a pn-junction operated in reverse bias.
Bias voltage is applied to the detector in order to achieve an electric field high enough for the charge carriers to undergo impact ionisation.
This mechanism is responsible for the multiplication of the charge carriers released by the passage of a charged particle or light impinging on the detector.

The pn-junction is located several tens of microns from the detector surface.
This, together with the shape of the doping profile, prevents the depletion region from reaching the detector's surface, avoiding breakdown and noise due to surface effects.
The concept of full depletion voltage, usually applied to detectors produced through the planar process, is not used with this type of APD.

\begin{figure}
  \centering
%  \subfloat{\includegraphics[width = 0.6 \columnwidth, trim=0.3cm 0.3cm 0.2cm 0.3cm, clip]{APD_Diagram}}\\
%  \subfloat{\includegraphics[width = 0.8 \columnwidth]{apdSectionNew}}
  \subfloat{\includegraphics[width = 0.3 \columnwidth, trim=0.3cm 0.3cm 0.2cm 0.3cm, clip]{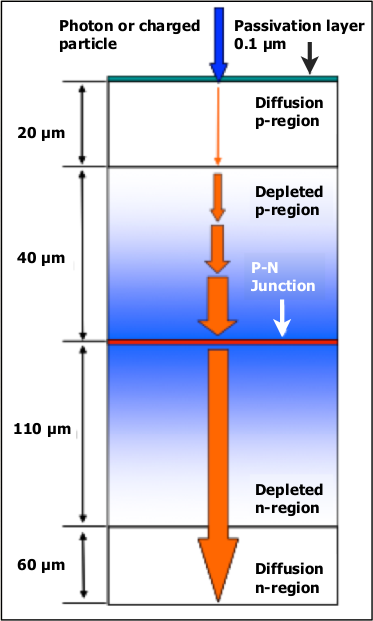}}\\
  \subfloat{\includegraphics[width = 0.4 \columnwidth]{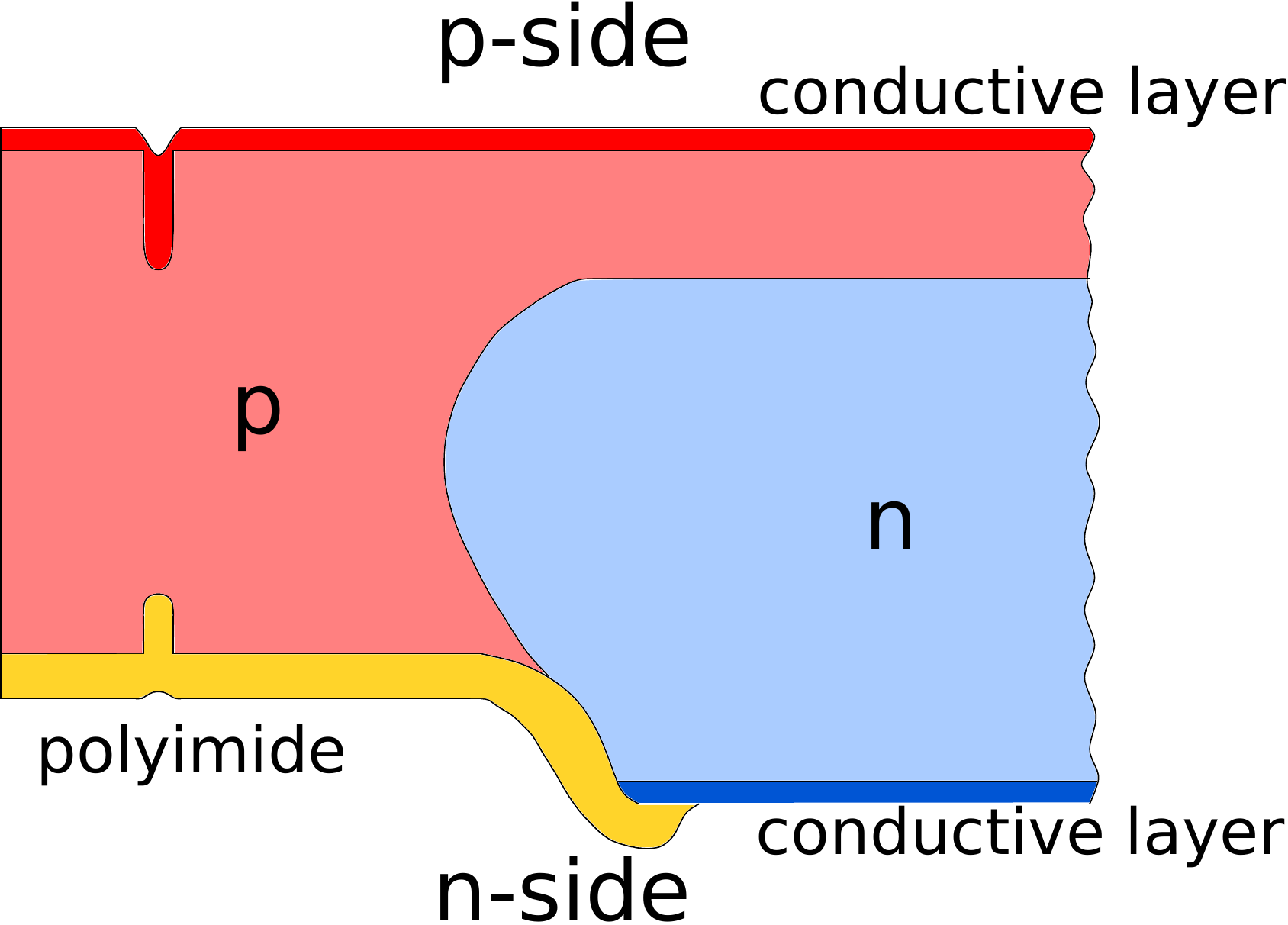}}
  \caption{Schematic cross-sections of a deep diffused APD. {\bf Top:} centre of the detector. The thickness of the depleted region corresponds to a bias voltage of 1.8\,kV. {\bf Bottom:} detector edge (colour online).}
  \label{fig:apdDia}
\end{figure}

The occurrence of breakdown and noise from the edges of the detector die is mitigated by reducing the electric field in these regions.
This can be achieved by bevelling the detector's edges\,\cite{mcclish2004}.
The current APD design reduces the electric field at the edges of the depletion region by distributing the doping in the sensor such that the depletion region is ``bent'' and terminated on one of the sensor's surfaces\,\cite{mcclish2004}.
A mesa structure, where the n-type silicon forms a plateau that is joined to the p-type silicon, reduces the electric field at the sensor's surface, see figure\,\ref{fig:apdDia}.

Deep diffused APDs get their name from the process used to produce a pn-junction several tens of microns from the detector surface.
The APDs used in this work are produced on silicon using the method explained in\,\cite{mcclish2004,apdPatent}.
Grooves are carved on both sides of an n-doped silicon wafer and p-type dopants are diffused into the silicon.
The grooves run parallel to the future die edges and shape the distribution of the p-type dopants into the silicon.
This process results in an n-doped region enclosed by the p-doped silicon.
The wafer is ground and polished on both sides.
One side is ground more than the other in order to reach the n-doped region.
At this point, the n-doped region is exposed on one side and surrounded by p-type silicon.
The pn-junction runs parallel to the detector surface and curves toward the side with the exposed n-doped region.
Non-metallic conductive layers are added on both sides to provide ohmic contacts to the p and n-doped volumes.
A mesa structure is then etched following the pn-junction on the side where the n-doped region is exposed.
Finally, polyimide is deposited around the mesa structure and the wafer is diced.

Schematic cross-sections of the centre and the edge of the resulting device are shown in figure\,\ref{fig:apdDia}.
In the following, the faces of the detectors will be referred to as p- and n-side, according to the sketch shown in figure\,\ref{fig:apdDia}.

The doping concentration of the n-type silicon is constant over the thickness of the detector, since it results from the initial doping of the silicon wafer.
The p-type silicon has a different doping profile.
As the p-type dopants are introduced through diffusion, the doping concentration of the p-type silicon is higher toward the p-side of the detector and falls toward the pn-junction.
This results in a broader peak of the electric field at the pn-junction, when compared to the triangular shape of an abrupt junction.
The broadening of the peak increases the distance over which the charge carriers drift in the highest field region, requiring a lower peak field for a given gain value.
%The broadening of peak allows the charge carriers to drift in the highest field region for a longer distance, requiring a lower peak field for a given gain value.
This reduces the ratio between the electron and hole multiplication coefficients, reducing in turn the excess noise factor due to multiplication\,\cite{theoryDDAPD}.

The applied bias voltage is usually around 1.8\,kV, resulting in a gain of up to\,500.
At this bias voltage the thickness of the depletion region is around 150\,$\mu$m.

\section{Samples and Irradiations}
\label{sec:samples}

All the APDs used in this study present the structure detailed in the previous section.
Detectors of two different sizes were used.
In this section, the geometry and readout of the different APDs are explained.

\subsection{$2 \times 2$~mm$^2$ APDs}

APDs with a nominal active area of $2 \times 2$~mm$^2$ were used to study the effects of neutron irradiation.
The dies of these devices have an area of $3.1 \times 3.1$~mm$^2$ and a circular mesa structure with a diameter of about 0.8\,mm.
These devices are mounted on ceramic supports, two metallic leads are used to contact the p- and n-sides of the detector.
The n-side of the detector faces the ceramic support.
The contact between the APD and the metal leads is achieved using conductive glue.

The $2 \times 2$~mm$^2$ APDs have a DC-coupled readout.
In order to facilitate the handling and electrical connection to the sensors, each APD, together with its ceramic support, was mounted on a printed circuit board (PCB) for its characterisation both before and after irradiation.
The PCB was equipped with a temperature sensor and the connectors needed to link the APD to the measuring devices.

The sensors were irradiated with neutrons at the nuclear reactor of the Jo\v{z}ef Stefan Institute in Ljubljana\,\cite{jsiIrrad}.
The fluences accumulated by the sensors ranged from $\Phi_{eq} = 3 \cdot 10^{13}$\,cm$^{-2}$ to $\Phi_{eq} = 10^{15}$\,cm$^{-2}$.
Neither bias nor cooling were applied during the irradiation, the temperature of the samples is estimated to range between 20 and 45$^\circ$C during irradiation\,\cite{vlado}.
After irradiation, the samples were stored at a temperature below $-18^\circ$C to avoid the annealing of the defects produced during irradiation.
No annealing steps were performed after irradiation, however, it is estimated that the annealing due to handling after irradiation corresponds roughly to one hour at room temperature.
%% However, the final total annealing due to the handling of the samples, and the performance of measurements at different temperatures, was estimated to be of 73 minutes at 21$^\circ$C. %% from Sofia's thesis

\subsection{$8 \times 8$~mm$^2$ APDs}

APDs with a nominal active area of $8 \times 8$~mm$^2$ were characterised in several beam tests.
The dies of these devices have an area of $10 \times 10$~mm$^2$ and a square mesa structure with 7.5\,mm sides.

As shown in section\,\ref{sec:unif8x8laser}, the amplitude of the signal produced by the sensor when illuminated by a laser depends on the distance between the point in which the sensor is struck by the laser and the point where the sensor is connected to the readout electronics.
This behaviour is the result of the non-negligible resistance of the conducting layers applied to the p- and n-side of the detector.

Two methods were used to improve the uniformity of response.
%% The first one, used in the devices characterised in beam tests, relies on an AC-coupled readout of the p-side of the detector.
The first one relies on an AC-coupled readout of the p-side of the detector.
A gold layer is applied to the n-side of the detector to improve its conductivity.
The electrical contact to the p-side is achieved through the formation of a bond-pad and wire-bonding.
The p-side is covered by a 50\,$\mu$m thick Kapton layer and a metallic mesh is placed above the Kapton.
The capacitance between the mesh and the APD is about 50\,pF.
The mesh picks-up the signal and provides an electrical path with a resistivity lower than the one of the conductive layer.
The sensors produced with this configuration were usually placed on PCBs containing the readout electronics.
These detectors are usually operated by applying the high voltage to the p-side and keeping the n-side at ground potential.
In this way it is possible for the electronics to directly read out both signals from the mesh and the n-side of the sensor.

Another method used to improve the uniformity of response is the metallisation of both the p- and n-side of the detector.
The metallisation consists in an aluminium deposition containing openings for laser illumination.
The p-side metallisation is continuous except for circular 2\,mm diameter hole at the detector centre.
The metallisation on the n-side consists of a grid structure with square openings with a side of 100\,$\mu$m.
The detectors produced with this configuration have a DC-coupled readout and were used in laboratory tests.
These sensors were mounted on the same kind of PCBs used for the $2 \times 2$~mm$^2$ APDs, that contain a temperature sensor and the connectors to bias and readout the sensor.

In this paper detectors with both DC- and AC-coupled readout are characterised, but at the time of writing only DC-coupled devices have been characterised after irradiation. 

\section{Gain Measurement of $8 \times 8$ mm$^2$ APDs using a Blue LED}
\label{sec:gain8x8led}

\begin{figure*}[h]
  \centering
  \subfloat{\includegraphics[width = 0.52 \textwidth, trim=0 0 0 1.95cm, clip]{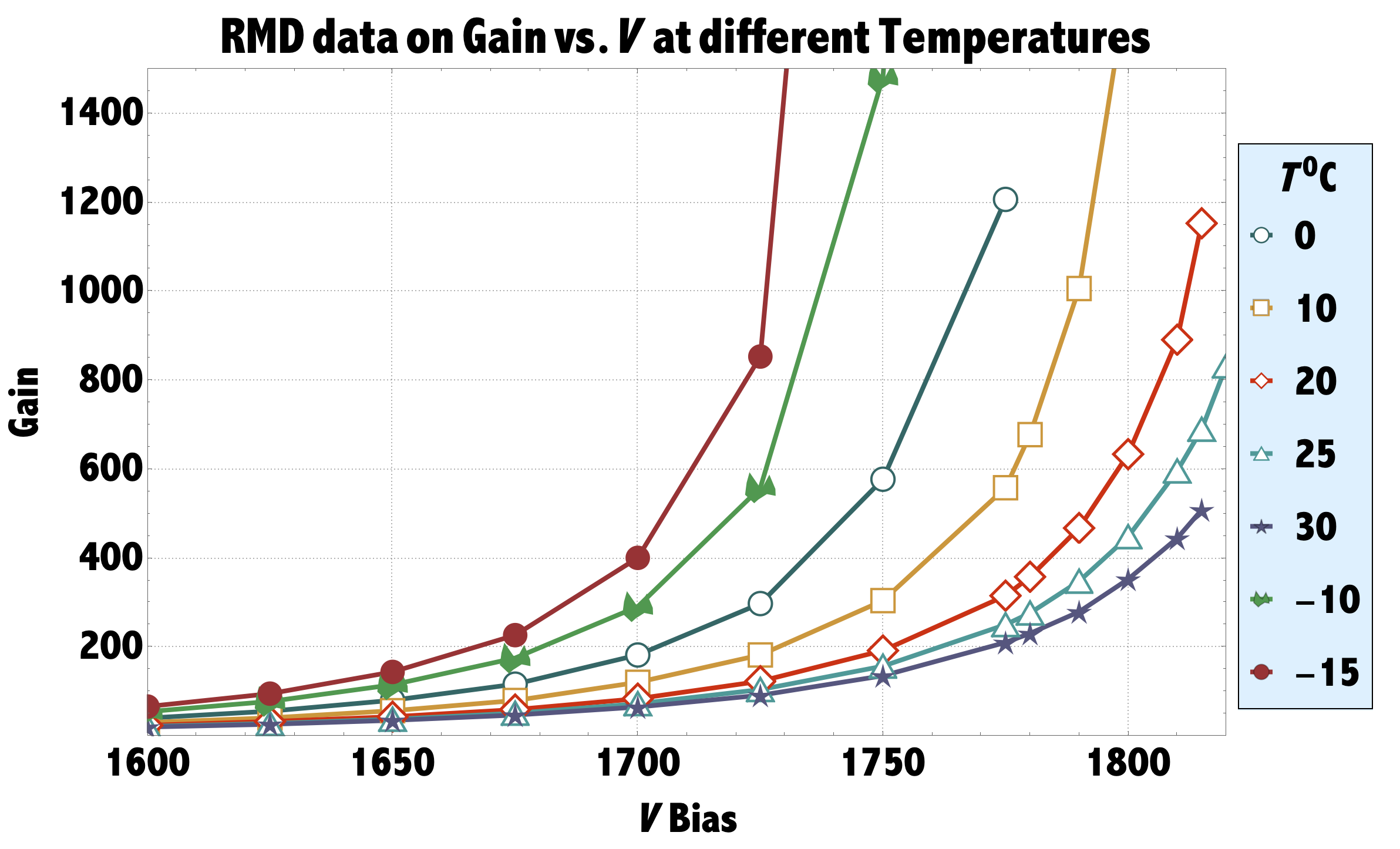}}
  \hfill
  \subfloat{\includegraphics[width = 0.48 \textwidth, trim=0 0 0 2.1cm, clip]{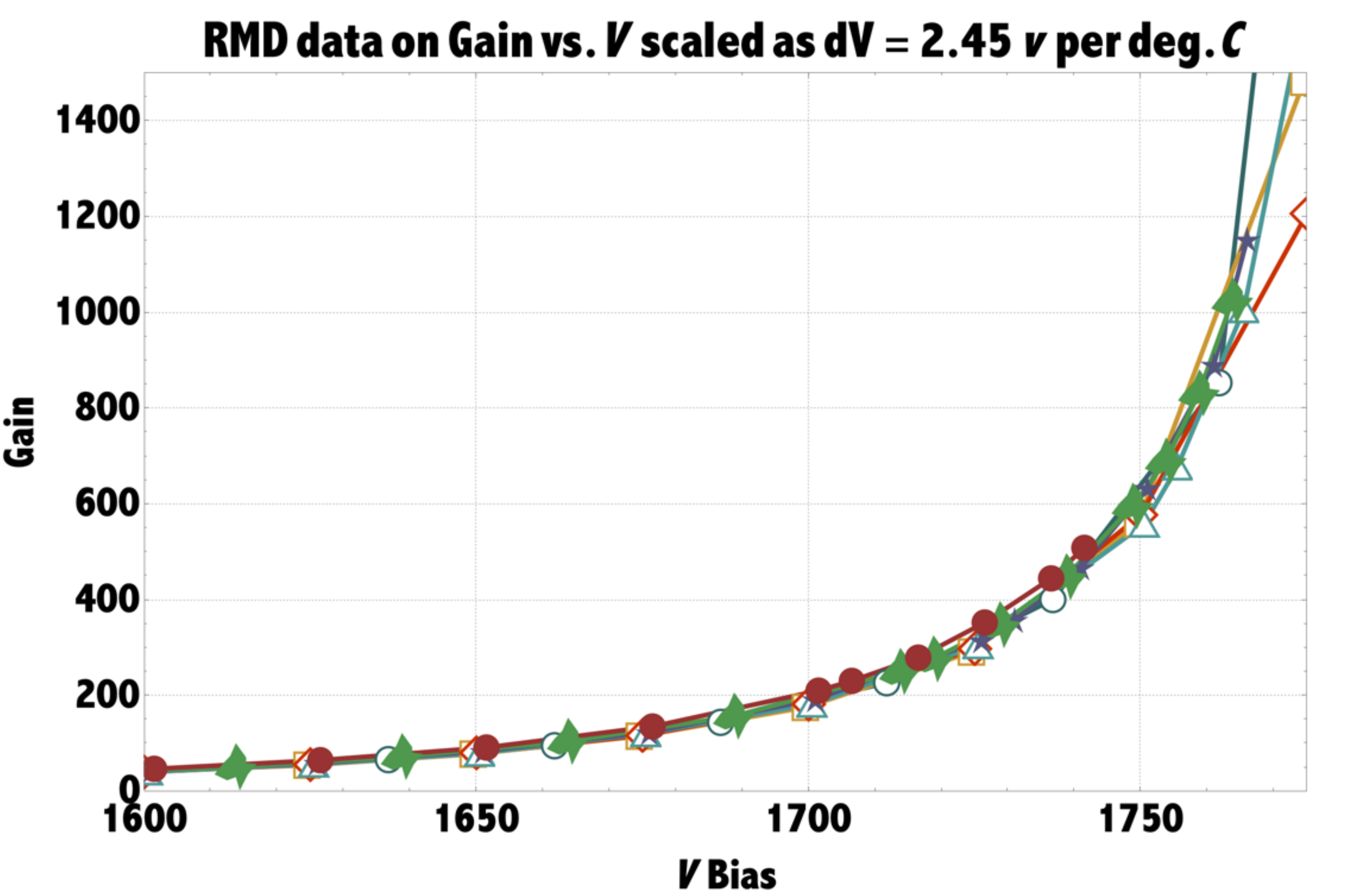}}
  \caption{\textbf{Left:} Gain of a $8 \times 8$ mm$^2$ APDs as a function of bias voltage at different temperatures. The measurement was performed using a blue LED. \textbf{Right:} By shifting the bias for each curve according to the law described in the text, a scaling law is found.}
  \label{fig:gainVsV}
\end{figure*}

The gain of $8 \times 8$ mm$^2$ APDs was measured using a pulsed LED.
The LED had a wavelength of 425\,nm and the pulses had a duration of 20\,ns.
The APD was uniformly illuminated during the measurements.
The readout chain consisted of a charge sensitive amplifier followed by a shaper with a 0.25\,$\mu$s time constant.
The signal was digitised using an oscilloscope, and the measured pulse height used for the determination of the gain.
Unity gain refers to the amplitude obtained at $\approx$500 V bias, where the amplitude shows a weak dependence on the bias voltage~$U$.
The measurements were performed at different temperatures, and the results are reported in figure\,\ref{fig:gainVsV}.
An attempt was made to parametrize the  voltage~$U$ required for a given gain as a function of temperature~$T$.
A temperature dependent shift $U' = U -2.45~\textrm{V}/^\circ\textrm{C}\cdot (T - T_0)$, with $T_0 = 0^\circ\textrm{C}$, describes the change in voltage necessary to normalise a given value of gain to the gain obtained at $0^\circ\textrm{C}$ when the temperature of the APD is changed.
%Here $\Delta T( 0^\circ\textrm{C})$ gives the temperature difference from $0^\circ\textrm{C}$. 
The result of this scaling law is shown in figure \ref{fig:gainVsV}(Right).
It describes the data up to a gain of $\approx$1000 and may prove useful for discriminating among impact ionisation models that appear in the literature and as options in TCAD\footnote{Technology Computer Aided Design.} software for modelling such devices.
It should be noted that this particular definition of gain, as it applies to visible light photodetection in an APD, is different from the gain obtained from the response to minimum ionising particles (or the IR laser model used) since, in the former case, all photoelectrons traverse the region where impact ionisation occurs.

\section{Laboratory Characterisation of Irradiated $2 \times 2$~mm$^2$ APDs}
\label{sec:irrad2x2}

\subsection{Experimental Methods}

Two experimental setups were used to characterise the APDs.

The current-voltage (IV) characteristic of the detectors was measured using a voltage source and a picoammeter connected in series to the sensor under test.
The sensor was placed inside a climate chamber flushed with dry air where the temperature was set to $-20^\circ$C and it could be controlled with an accuracy of $0.2^\circ$C.
The temperature sensor on the PCB was used to ensure that the APD reached thermal equilibrium with the air in the climate chamber before starting the measurements.

The response of the APDs to a pulsed infrared laser was measured in a different setup and used to determine the variation of the sensor's signal and time resolution as a function of bias voltage and irradiation fluence.
The laser has a wavelength of 1064\,nm and the pulses have a duration of 200\,ps.
The repetition rate was chosen to be 200\,Hz.
The intensity of the light impinging on the APD was determined to correspond to a charge deposition in the sensor of 15 or 0.8\,MIPs per pulse, depending on the measurement.
A variable slit is used to change the light intensity.
The amount of deposited charge per laser pulse was measured using a non-irradiated pad diode of known thickness.
The pad diode signal was digitised using an oscilloscope and integrated, leading to the deposited charge.
These measurements were repeated with and without the use of an amplifier and yield to the same result.
The charge deposited for one MIP is defined in this work as 74 electron hole pairs per micron of silicon.
The wavelength used has an absorption length in silicon of about 1\,mm, resulting in the generation of electron hole pairs through the whole sensor thickness.
The light pulses are propagated from the laser to the sensor through an optical fibre.
A coupler diverts part of the light to a photodiode that is used to monitor the intensity of the light pulses.
The relation between the photodiode response and the amount of charge produced by the light pulses was determined using the pad diode measurements mentioned above.
An optical system focuses the light on the sensor, producing a beam spot with a full width half maximum of about 35\,$\mu$m as measured on the sensor surface using a knife-edge technique.
The sensor is mounted on a 3 axis linear stage system enabling movement of the sensor with respect to the laser focusing system.
The bias voltage of the sensor under test is provided by a voltage source containing a picoammeter used to monitor the current flowing through the sensor.
The temperature of the APD was controlled using a cooling system consisting of a Peltier element and a chiller.
The temperature measured by the temperature sensor on the PCB was used as input to the Peltier control system.
The APD was housed in a light-tight Faraday cage flushed with dry air during the measurements.
The APD signal was amplified using a CIVIDEC C2HV broadband amplifier\,\cite{cividec}.
Both the signals of the APD and the photodiode were digitised using an oscilloscope with 2.5\,GHz bandwidth and a sampling rate of 20\,GSa/s.

For the measurement of the APD signal amplitude and uniformity of response, an amplification of 10\,dB was used, together with a light intensity corresponding to 15\,MIPs.
This amplification was achieved by attenuating the APD signal with a 30\,dB attenuator and then amplifying it with a 40\,dB amplifier.
The 40\,dB amplifier is the same used in the time resolution measurements.
The reduction of the gain to 10\,dB is necessary in order to perform the amplitude measurements over the desired bias voltage range while remaining in the amplifier's linear range.
For each measurement condition, the waveforms were averaged 256 times in the oscilloscope before being stored for analysis.

The time resolution measurements were performed applying a 40\,dB amplification to the APD signal.
No averaging was applied to the waveforms.
A gain of 40\,dB provides a better signal to noise ratio for the APD signal, compared to the 10\,dB used for the amplitude measurement.
The intensity of the light shone on the APDs for these measurements corresponded to 0.8\,MIPs.
The optical system was modified for the timing measurements.
A splitter and delay line system was introduced, thereby projecting two light pulses on the sensor for each pulse generated by the laser.
The system is realised using optical fibre, and consisted of a splitter, a short and a long optical fibre branch, and a merger.
The difference in length between the two branches corresponds to a delay of 50\,ns between the pulses.
The sensor's signals from the light pulses were digitised in the same waveform in the oscilloscope.
The difference in the amplitude of the two signals was below 5\%.
Given the similar amplitude, the signals can be used to determine the time resolution of the detector under test, without the need of an external timing reference.

The APD temperature was $-20^\circ$C during all measurements reported in this section.
For the measurements of intensity as a function of bias voltage and the timing measurements using the laser, the light was shone on the centre of the APDs.

\subsection{Results}

\begin{figure}
  \centering
  \includegraphics[width = 0.6 \columnwidth]{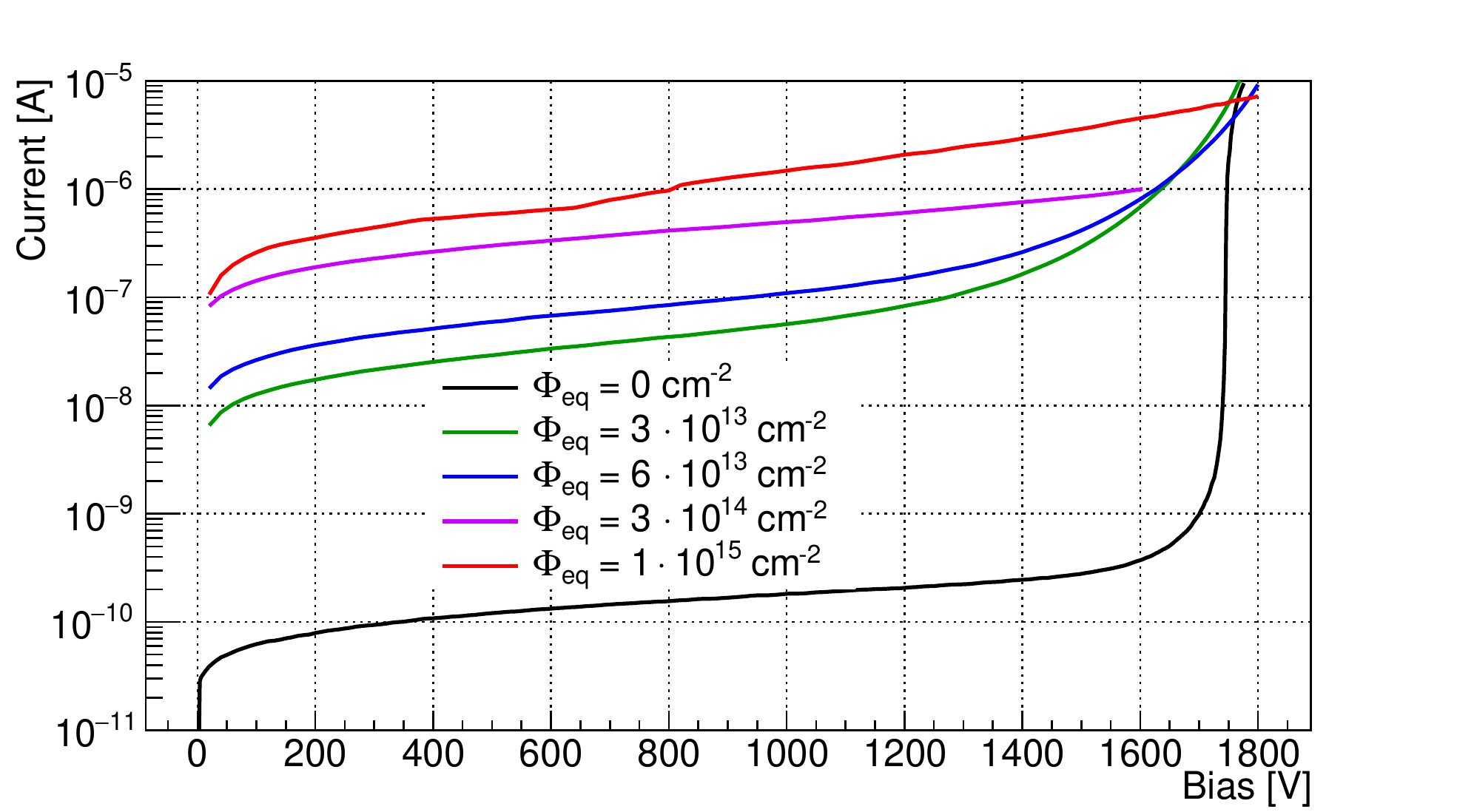}
  \caption{Current-voltage characteristics of the $2 \times 2$~mm$^2$ APDs measured at $-20^\circ$C for different neutron fluences (colour online).}
  \label{fig:iv2x2}
\end{figure}

\paragraph{Current-voltage characteristic}
The current-voltage (IV) characteristics of the APDs are shown, for different fluences, in figure\,\ref{fig:iv2x2}.
Before irradiation, below 1600\,V, the current assumes a value of less than a few nA.
In this region the main contribution to the current is thought to be due to surface current.
Between 1600 and 1800\,V, the current increases by 4 orders of magnitude.
This is the region where the multiplication dominates the IV characteristic of the non-irradiated sensors.
The surface current is not affected by multiplication, therefore the multiplication of bulk current in the non-irradiated sensors can not be seen in the IV curve until the gain is sufficiently high.

The irradiation enhances the bulk generation current of the devices, as can be seen in the region between 0 and 1200\,V, where the gain of the sensors does not influence the curves.
As the bulk current is amplified, the shape of the irradiated sensors' curves is different from that of the non-irradiated sensor.
The change in the magnitude of the current at high voltages is different with respect to the non-irradiated sensor, suggesting that the gain of the detectors is reduced by irradiation, or respectively that a higher bias voltage is required to obtain the same gain as for the non-irradiated sensor.

The current related damage rate ($\alpha$) was estimated at 200\,V, where no multiplication is expected.
Under the assumption that the depleted volume at this voltage does not change with irradiation, the measured value of $\alpha$ is of the expected order of magnitude\,\cite{ugobonoThesis}.
Further studies of the current generation in these detectors after irradiation can also be found in\,\cite{ugobonoThesis}.

The sensor irradiated to $\Phi_{eq} = 3 \cdot 10^{14}$\,cm$^{-2}$ shows a breakdown around 1600\,V, therefore no information about its IV characteristic is available at higher bias voltages.
This also applies to the amplitude measurements presented below.

\begin{figure*}[h]
  \centering
  \subfloat[$\Phi_{eq} = 0$\,cm$^{-2}$, 1700\,V]{\includegraphics[width = 0.49 \textwidth]{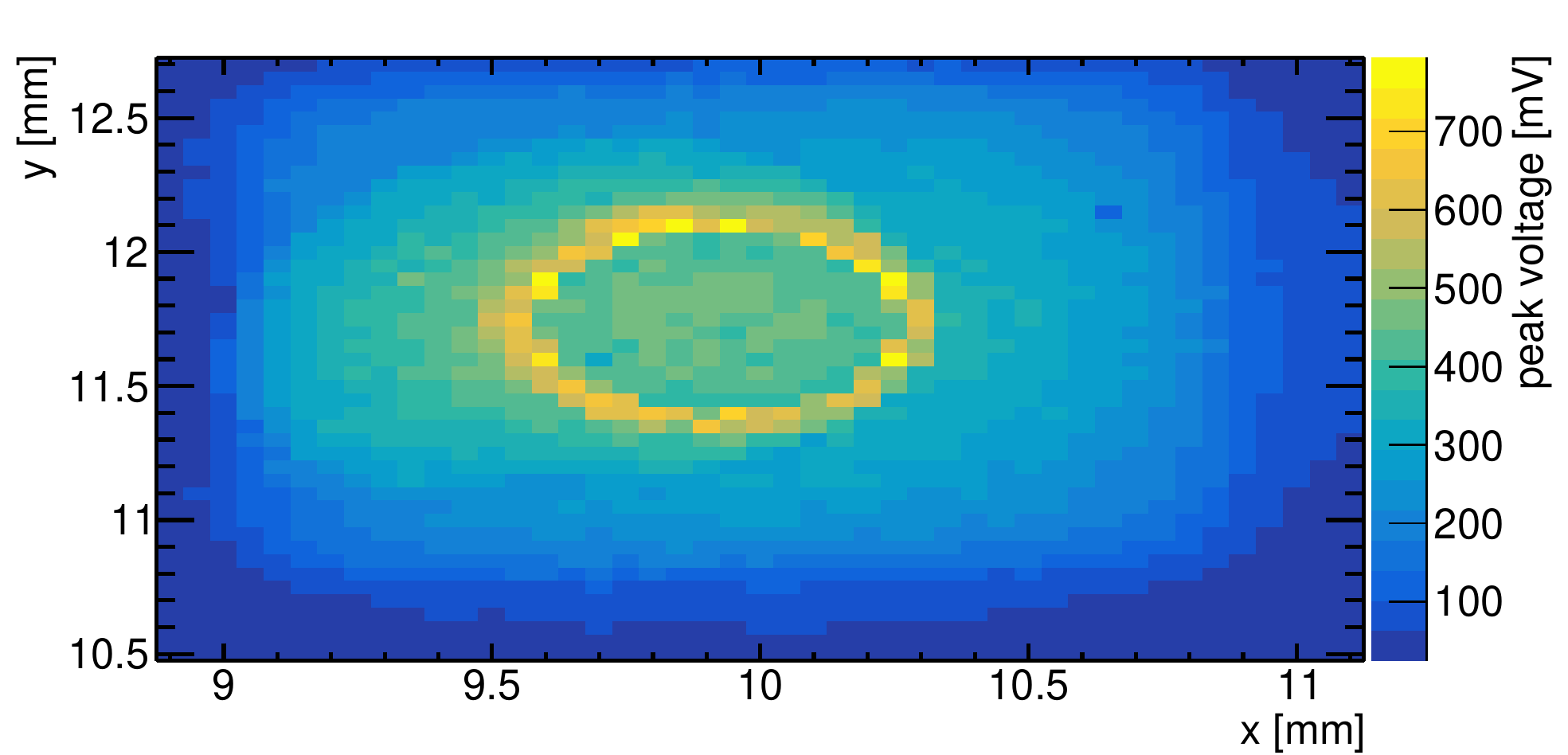}}\\
  \subfloat[$\Phi_{eq} = 3 \cdot 10^{13}$\,cm$^{-2}$, 1679\,V]{\includegraphics[width = 0.49 \textwidth]{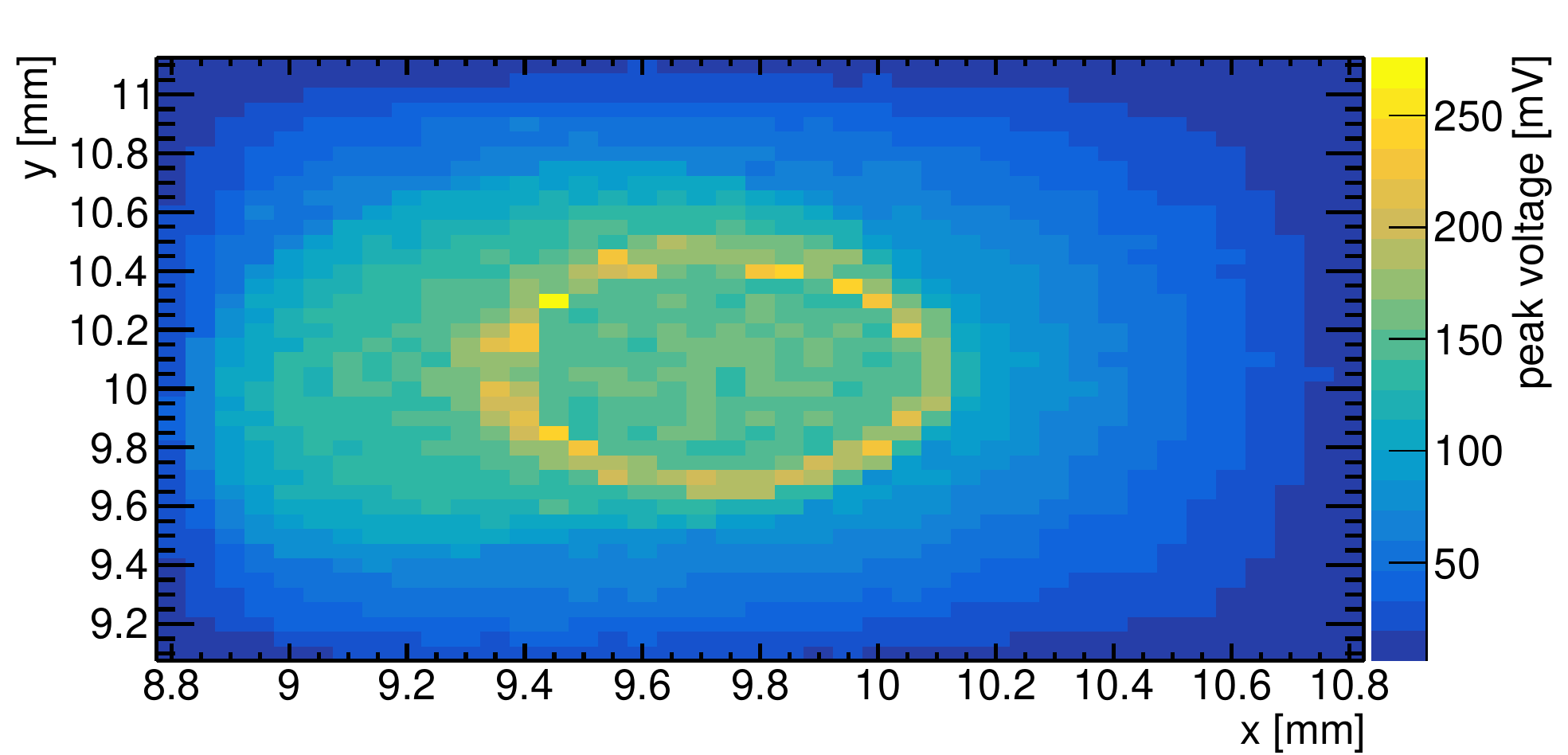}}\hfill
  \subfloat[$\Phi_{eq} = 6 \cdot 10^{13}$\,cm$^{-2}$, 1673\,V]{\includegraphics[width = 0.49 \textwidth]{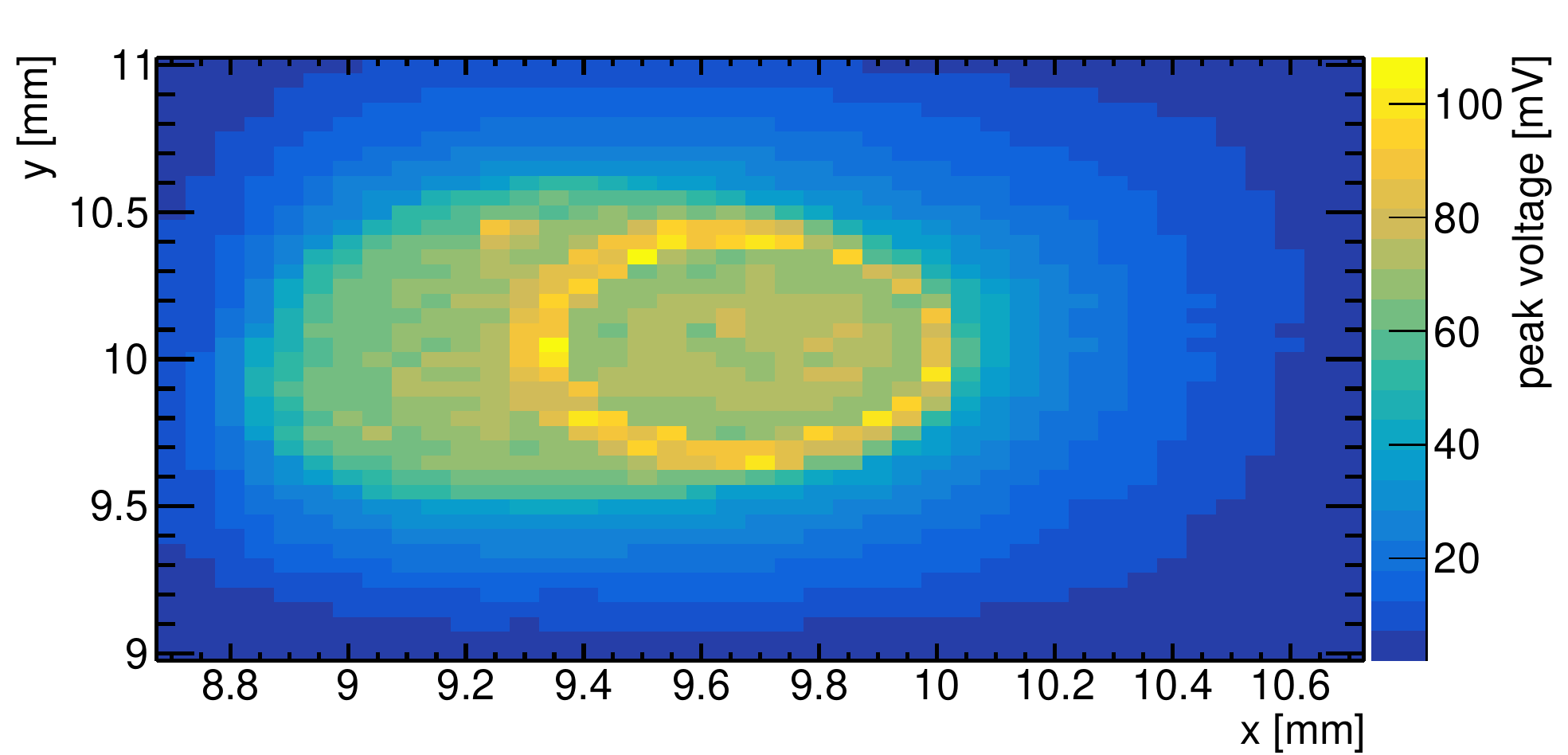}}\\
  \subfloat[$\Phi_{eq} = 3 \cdot 10^{14}$\,cm$^{-2}$, 1677\,V]{\includegraphics[width = 0.49 \textwidth]{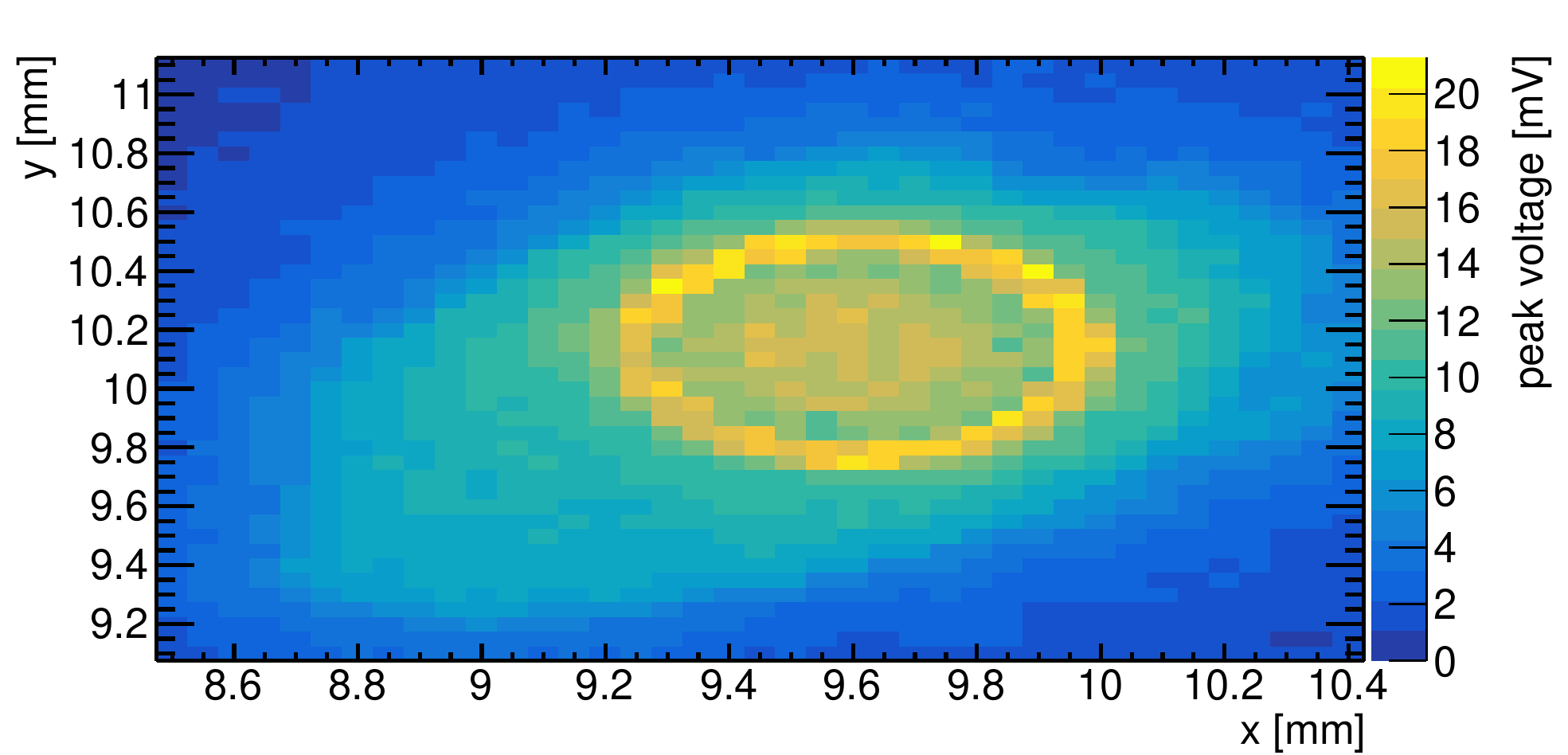}}\hfill
  \subfloat[$\Phi_{eq} = 10^{15}$\,cm$^{-2}$, 1644\,V]{\includegraphics[width = 0.49 \textwidth]{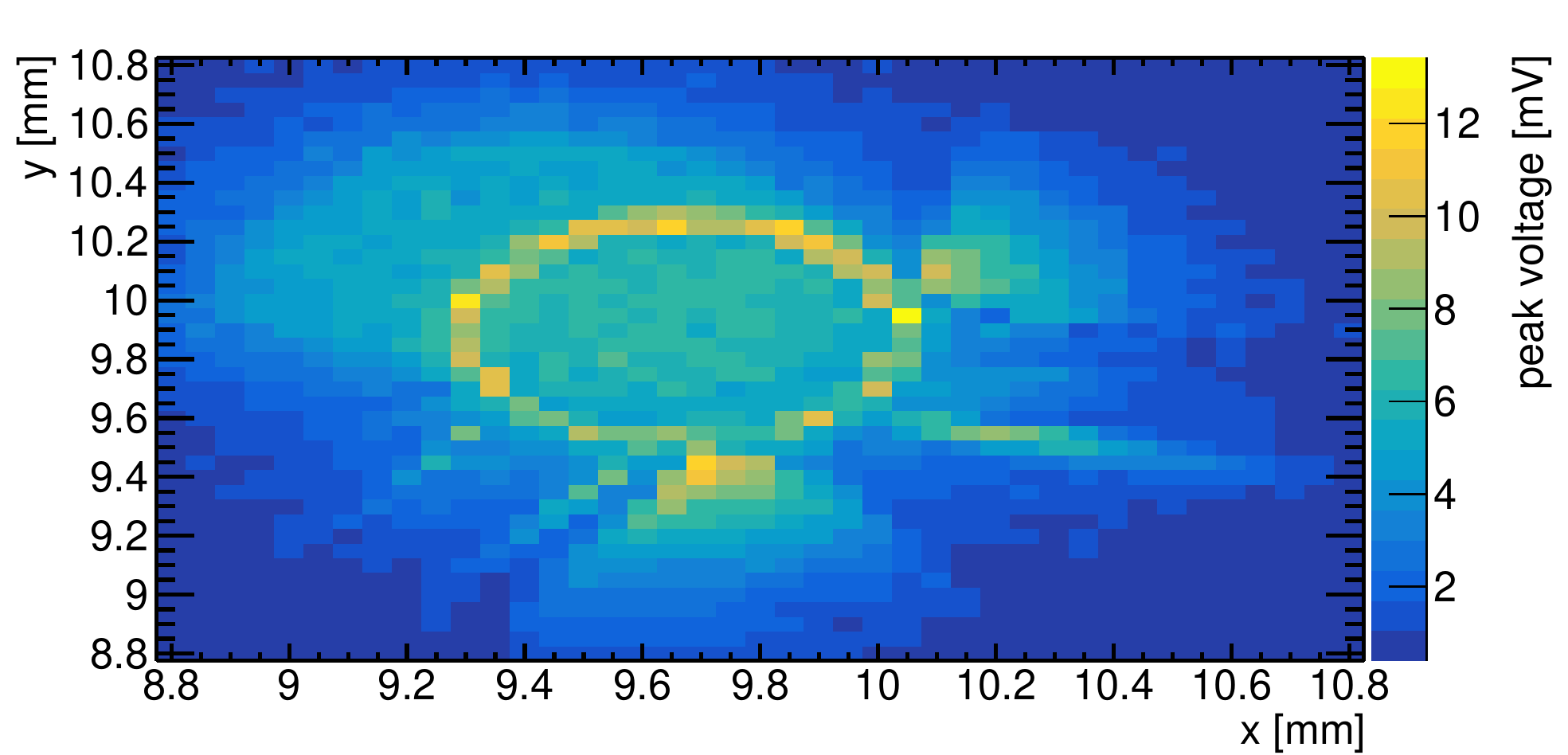}}
  \caption{Signal amplitude as a function of the $x-y$ position where the laser was focused on the sensors measured at $-20^\circ$C. The laser intensity corresponds to 15\,MIPs. An amplification of 10\,dB was used. The colour scale was adjusted for each fluence (colour online).}
  \label{fig:xyScanIR_2x2}
\end{figure*}

\paragraph{Uniformity of response}
The uniformity of response was measured using an infrared laser with pulses of an intensity corresponding to 15\,MIPs and an amplification of 10\,dB.
The laser was focused on different parts of the sensors in a 50\,$\mu$m step grid and the average of 256 waveforms was stored and the signal amplitude extracted for each position.
The signal amplitude as a function of position of the laser spot on the sensors is shown in figure\,\ref{fig:xyScanIR_2x2}.
The difference in the bias voltage applied to the detectors is due to the current drawn by the detectors under bias.
The biasing and readout circuit had a total resistance of 13\,M$\Omega$ connected in series to the sensor under test.
The bias voltage applied to the readout circuit was 1700\,V for all samples, the difference in voltage from this value are due to the potential drop caused by the sensors' dark current flowing through the 13\,M$\Omega$ load.
All the values of bias voltage presented in the following take into account this effect.

The sensor irradiated to $\Phi_{eq} = 3 \cdot 10^{14}$\,cm$^{-2}$ could be biased at a higher voltage with respect to the IV measurement.
This is probably a consequence of the annealing of the radiation damage of the detector, due to handling and measurements performed at room temperature, since the measurement shown in figure\,\ref{fig:xyScanIR_2x2} were performed after the current-voltage and amplitude measurements.
This suggests that the annealing status of the detectors can influence their breakdown behaviour.

The elongated ovals in figure\,\ref{fig:xyScanIR_2x2} indicate a larger signal amplitude in these regions.
This effect, although not fully understood for these devices, is thought to be similar in origin to the one studied in section\,\ref{sec:unif8x8laser} for APDs with an active area of $8 \times 8$~mm$^2$.
In each plot of figure\,\ref{fig:xyScanIR_2x2}, a circular region with increased amplitude is present.
This region corresponds to the mesa structure in the back of the devices.
Part of the light from the laser is reflected from the curved surface of the mesa toward the active area of the detectors, resulting in an increased signal.
The sensitive area of the detectors is affected by irradiation.
However, the response of the detectors appears uniform close to the detector centre, in the area enclosed by the mesa structure.
The measurements shown in the rest of this section were performed by shining the light in the centre of the detectors.

\begin{figure}
  \centering
  \includegraphics[width = 0.6 \columnwidth]{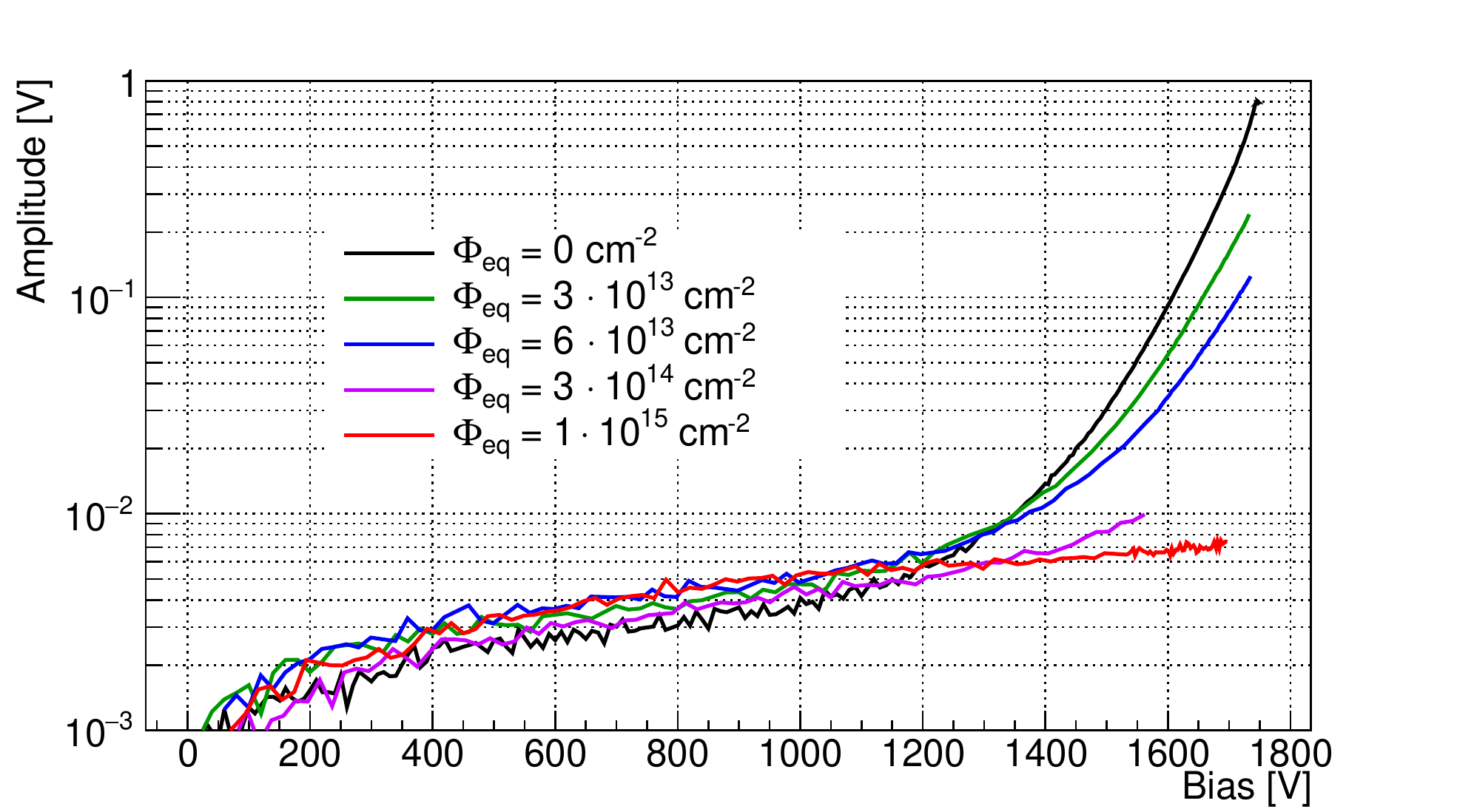}
  \caption{Amplitude of the APDs signal as a function of bias voltage and fluence measured at $-20^\circ$C. The laser intensity corresponds to 15\,MIPs. An amplification of 10\,dB was used (colour online).}
  \label{fig:ampli2x2_15MIP}
\end{figure}

\paragraph{Signal amplitude}
The amplitude was measured from the baseline of the pulse, calculated using the part of the waveform preceding the laser pulse.
The amplitude was determined using a parabolic interpolation of the waveform highest point and its first neighbours.
The results are shown in figure\,\ref{fig:ampli2x2_15MIP} as a function of bias voltage and fluence.

The measured amplitudes present a spread of less than 3\,mV up to 1200\,V bias, for all the fluences.
For voltage values above 1200\,V, the signal amplitude decreases with increasing fluence, indicating a reduction of the gain for the irradiated detectors.
The behaviour of the detectors indicates that for fluences at least up to $\Phi_{eq} = 6 \cdot 10^{13}$\,cm$^{-2}$ the gain of the detectors could be recovered by applying a higher bias voltage.

A similar gain degradation was observed for LGADs and attributed to a reduction of the doping concentration of the p-doped gain layer of these detectors with irradiation, the so-called acceptor removal process\,\cite{kramberger2015}.
A study linking this effect to the defects created in the silicon lattice by irradiation can be found in\,\cite{gurimskaya2019}.
The description of the reduction of the gain of the APDs with irradiation is beyond the scope of this paper.

\begin{figure}
  \centering
  \includegraphics[width = 0.6 \columnwidth]{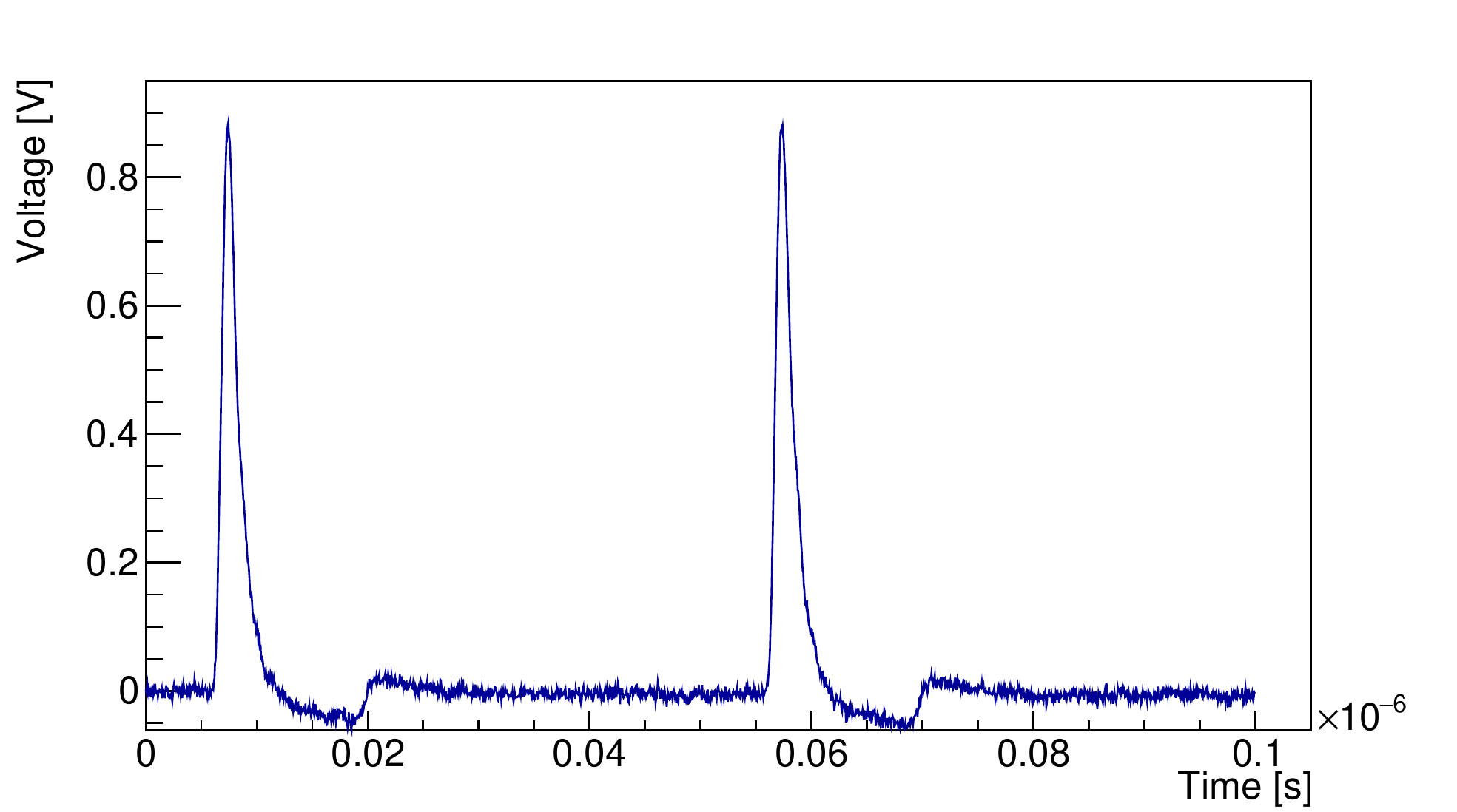}
  \caption{Waveform from a non-irradiated $2 \times 2$~mm$^2$ APD biased at 1665\,V operated at $-20^\circ$C. The laser intensity corresponds to 0.8\,MIPs. An amplification of 40\,dB was used. The signal undershoot is caused by the readout electronics.}
  \label{fig:pulses2x2timing}
\end{figure}

\paragraph{Time resolution}
The time resolution of the sensors was determined using an infrared laser focused on the detector centre.
An optical system was used to have two light pulses shine on the sensor under test for each pulse produced by the laser.
A waveform corresponding to one laser pulse is shown in figure\,\ref{fig:pulses2x2timing}.
The amplification used for the time resolution measurements was 40\,dB, and no averaging was applied to the waveforms.
2000 waveforms were acquired for each measurement condition.

\begin{figure}
  \centering
  \includegraphics[width = 0.6 \columnwidth]{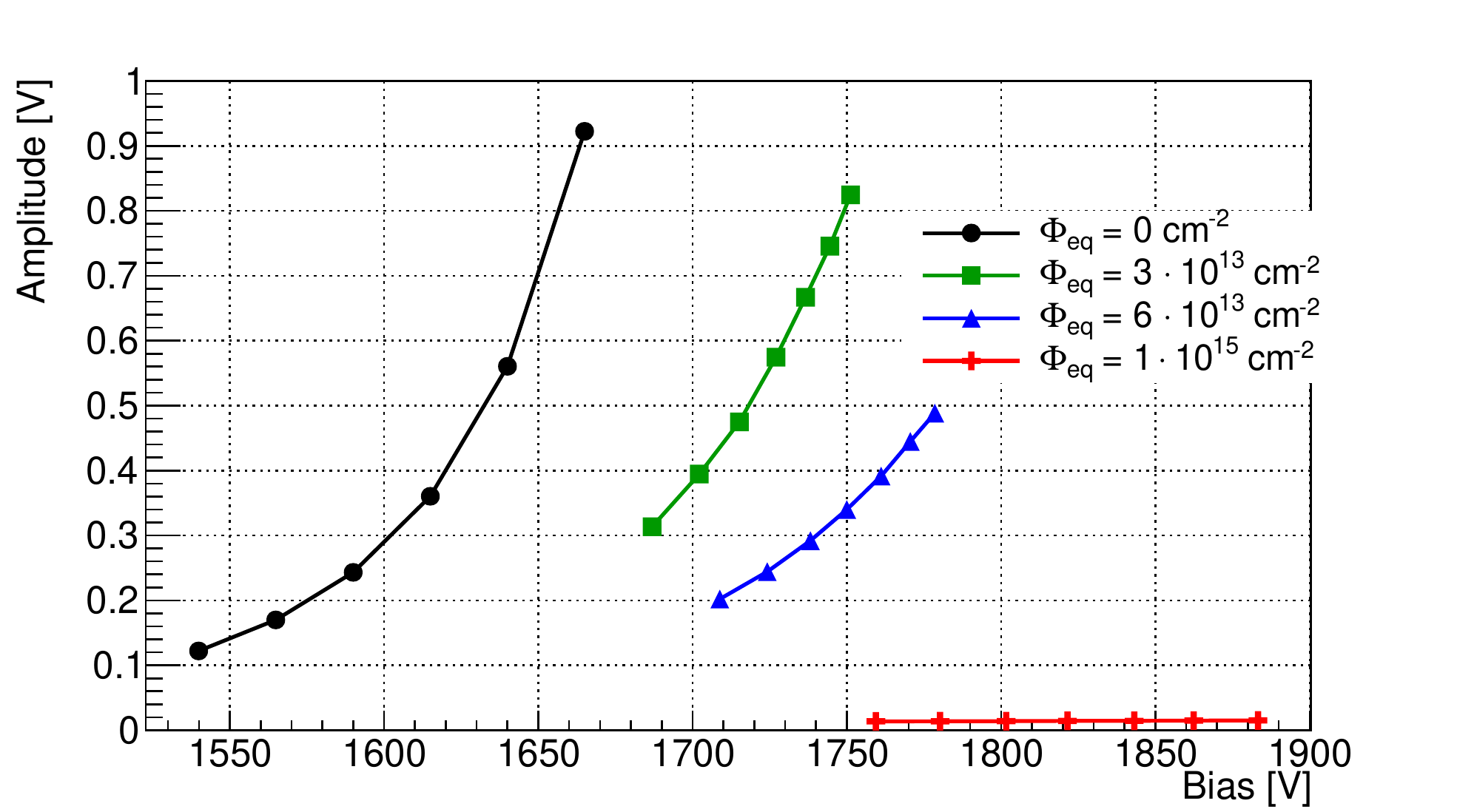}
  \caption{Average amplitude of the $2 \times 2$~mm$^2$ APDs signal as a function of bias voltage and fluence measured at $-20^\circ$C. The laser intensity corresponds to 0.8\,MIPs. An amplification of 40\,dB was used.}
  \label{fig:ampli2x2}
\end{figure}

The average signal amplitude as a function of bias voltage and fluence is shown in figure\,\ref{fig:ampli2x2}.
The amplitude is limited by two components of the experimental setup.
The limiting factor for the non-irradiated sensor is the amplifier, that has a linear range corresponding to an output amplitude of $\pm 1$\,V.
For the irradiated sensors, the maximum current provided by the high voltage power supply was reached, limiting the gain achieved by the devices.
The difference between the amplitude of the two signal peaks present in each waveform was less than 5\% for each measurement.
The time resolution of the sensor irradiated to $\Phi_{eq} = 3 \cdot 10^{14}$\,cm$^{-2}$ was not measured due to its breakdown behaviour that poses a risk to the electronics used in these measurements.

\begin{figure}
  \centering
  \includegraphics[width = 0.6 \columnwidth]{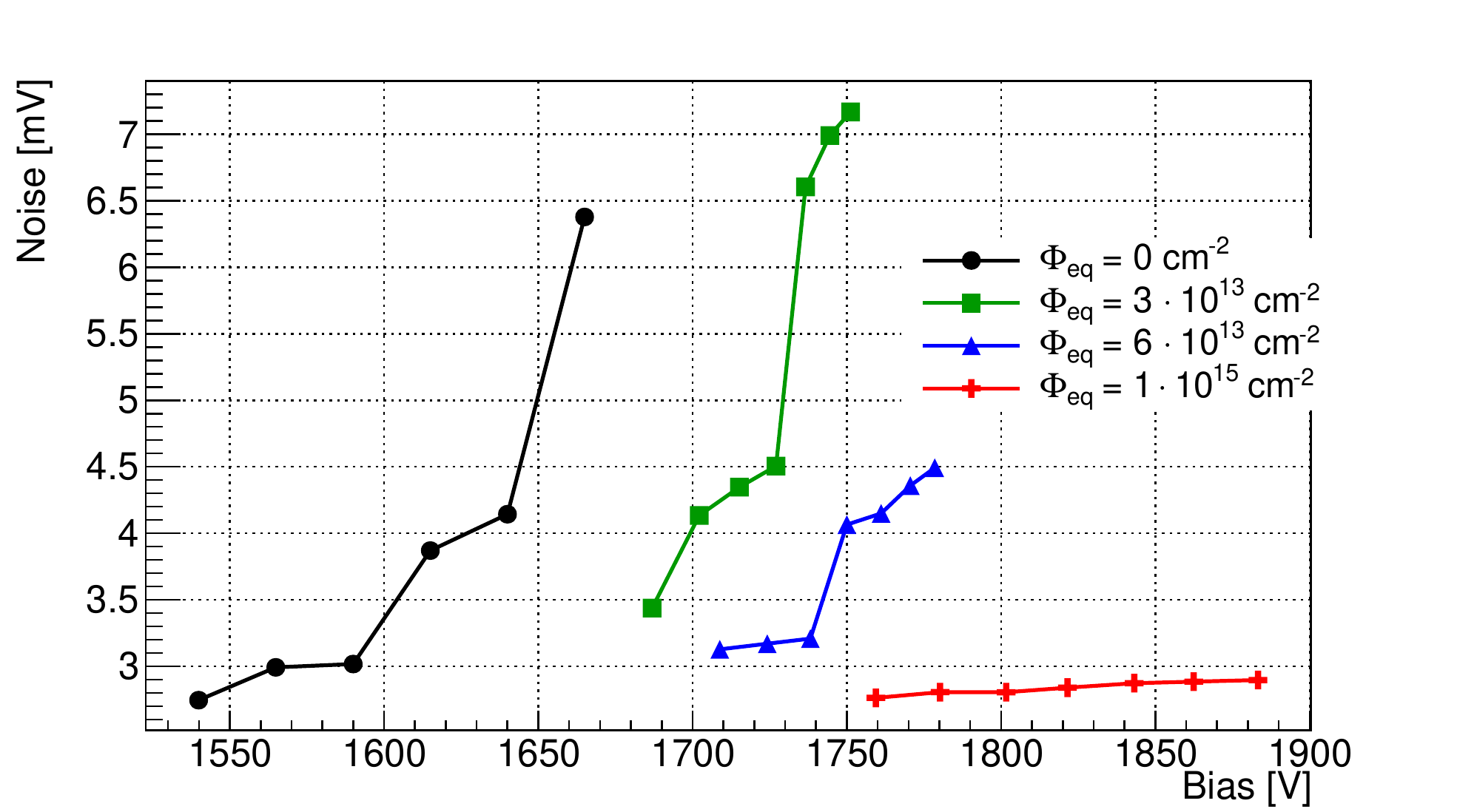}
  \caption{$2 \times 2$~mm$^2$ APDs' noise as a function of bias voltage and fluence measured at $-20^\circ$C. An amplification of 40\,dB was used.}
  \label{fig:noise2x2}
\end{figure}

The noise is defined as the standard deviation of the distribution of the waveform points around the baseline, for a 5\,ns portion of the waveform preceding the pulse due to laser illumination.
The noise of the APDs, as a function of bias voltage and irradiation fluence, is shown in figure\,\ref{fig:noise2x2}.
The sharp increases in the noise of the sensors irradiated to $3$ and $6 \cdot 10^{13}$\,cm$^{-2}$ between 1700 and 1750\,V are the result of the automatic adjustment of the oscilloscope vertical scale to accommodate the increase of signal amplitude with voltage.
The increase in the vertical scale of the oscilloscope results in an increase of the measured noise (i.e.\ since the least count of the measuring system contributes digital noise).
A similar effect can be seen for the non-irradiated sensor around 1650\,V.
Even accounting for the effects of the oscilloscope scale adjustment, the noise of the APDs with $\Phi_{eq} \leq 6 \cdot 10^{13}$\,cm$^{-2}$ shows a stronger dependence on the applied bias voltage than the the detector irradiated to $\Phi_{eq} = 10^{15}$\,cm$^{-2}$.
This suggests that the multiplication mechanism affects the noise of the sensors.

\begin{figure}
  \centering
  \includegraphics[width = 0.6 \columnwidth]{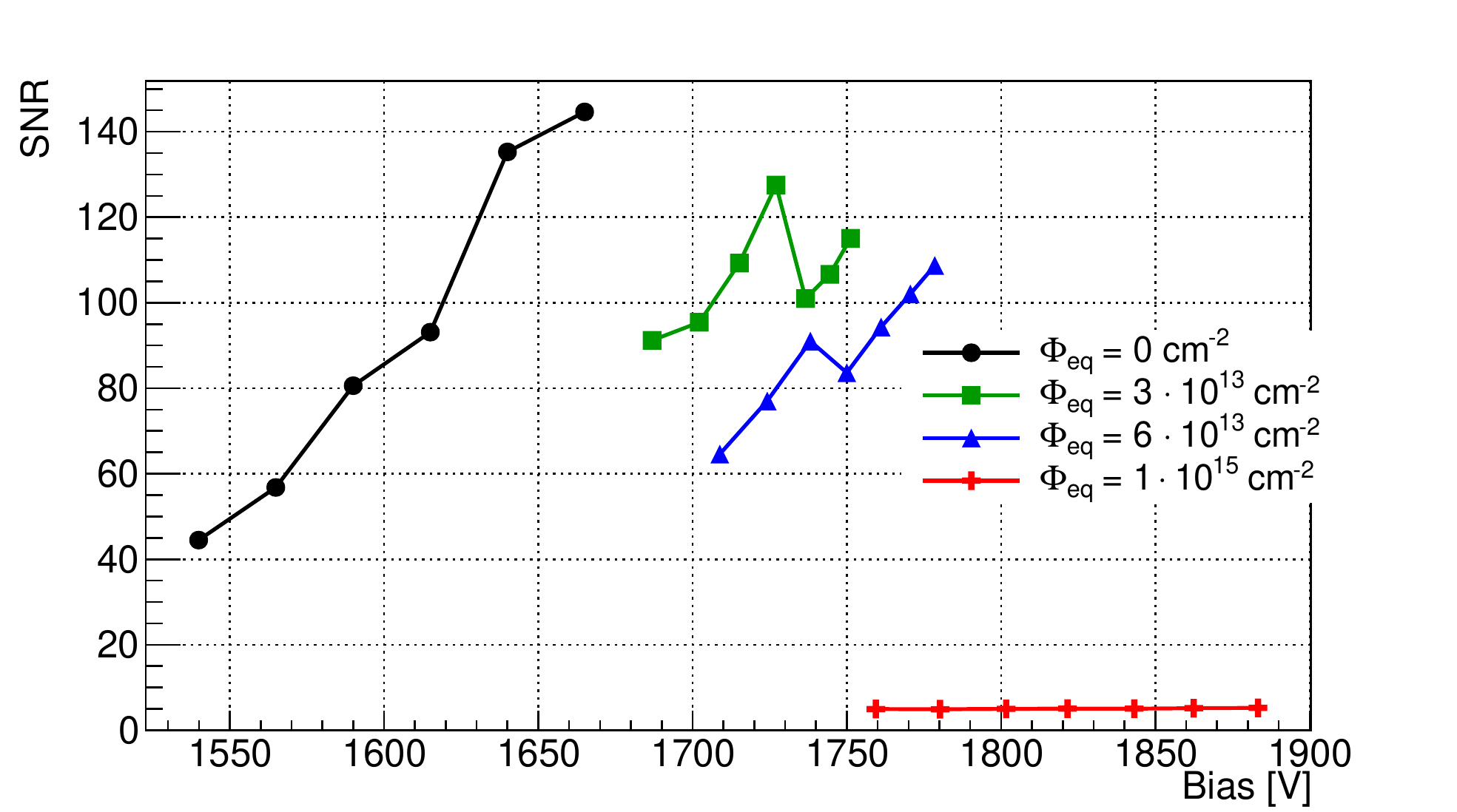}
  \caption{Signal to noise ratio (SNR) of the $2 \times 2$~mm$^2$ APDs as a function of bias voltage and fluence measured at $-20^\circ$C. The laser intensity corresponds to 0.8\,MIPs. An amplification of 40\,dB was used.}
  \label{fig:snr2x2}
\end{figure}

The signal to noise ratio (SNR) is defined as the ratio between amplitude and noise and is shown in figure\,\ref{fig:snr2x2}.
The SNR is not a monotonic function of the applied bias voltage.
This is the result of the adjustments of the oscilloscope vertical scale discussed above.
In contrast to the other sensors, the sensor irradiated to $\Phi_{eq} = 10^{15}$\,cm$^{-2}$ shows no dependence of SNR on bias voltage.
This is attributed to the lower value of gain of this sensor compared to the others.

\begin{figure}
  \centering
  \includegraphics[width = 0.6 \columnwidth]{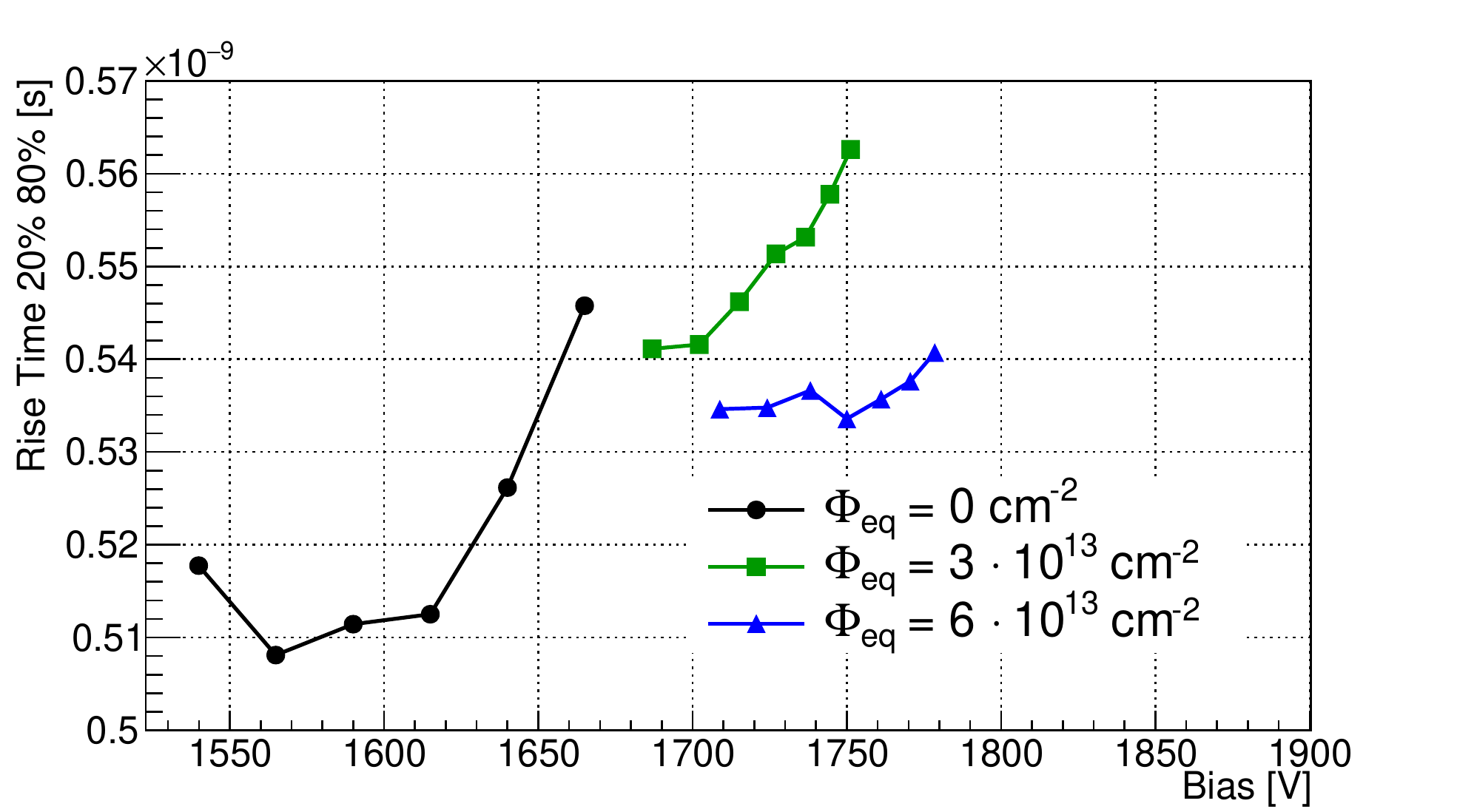}
  \caption{20\%-to-80\% rise time of the $2 \times 2$~mm$^2$ APDs signal as a function of bias voltage and fluence measured at $-20^\circ$C. The laser intensity corresponds to 0.8\,MIPs. An amplification of 40\,dB was used.}
  \label{fig:riseTime2x2}
\end{figure}

The average 20\%-to-80\% rise time of the sensors is shown in figure\,\ref{fig:riseTime2x2}.
The crossing time of the 20\% and 80\% thresholds was determined using a linear interpolation between two points of the waveform.
The values for the sensor irradiated to $\Phi_{eq} = 10^{15}$\,cm$^{-2}$ lie between 5.1 and 5.3\,ns and are not shown since they are affected by its low SNR, that does not allow for a correct determination of the 20\% point.
The rise time increases with increasing bias voltage.
% how to explain this?
% higher V -> higher depleted thickness -> lower weighting field?
% higher V -> higher depleted thickness -> more space for the last electrons in the p-region to reach the multiplication region? the n-side has a higher depleted thickness, so hole collection does not matter?
The difference between the rise time of the sensors lies within 11\%, the rise time does not seem to be affected by irradiation for $\Phi_{eq} \leq 6 \cdot 10^{13}$\,cm$^{-2}$.

%1e15 -> risetime 5.1-5.3 ns

\begin{figure}
  \centering
  \includegraphics[width = 0.6 \columnwidth]{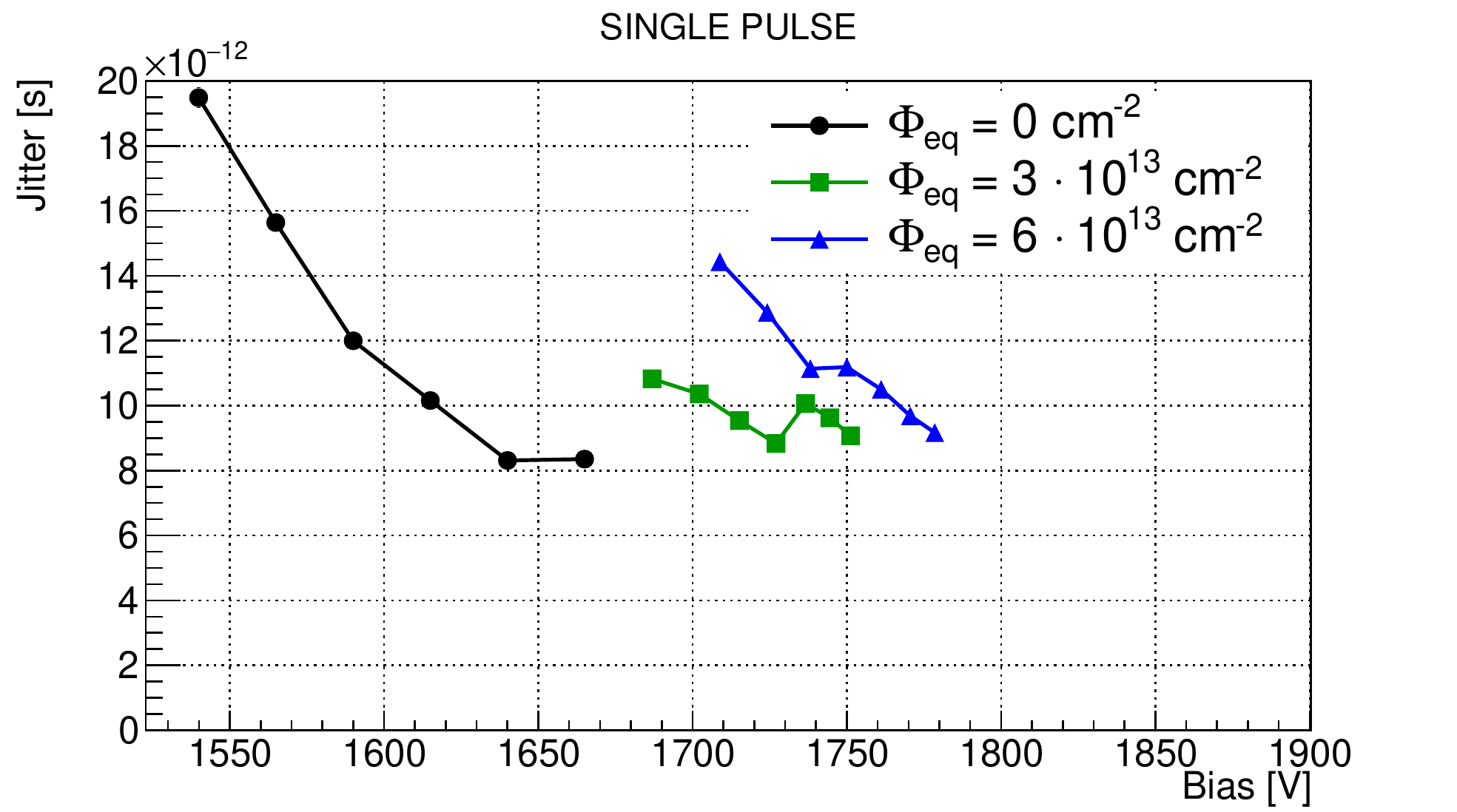}
  \caption{Single pulse time resolution of the $2 \times 2$~mm$^2$ APDs as a function of bias voltage and fluence measured at $-20^\circ$C. The laser intensity corresponds to 0.8\,MIPs. An amplification of 40\,dB was used.}
  \label{fig:timeRes2x2}
\end{figure}

The time resolution of the sensors was determined using the two pulses acquired in each waveform.
The waveforms are divided in two parts of 50\,ns each.
The pulses are analysed independently.
The time difference between the pulses is calculated using a constant fraction discriminator (CFD) algorithm.
This algorithm was chosen since it allows to study the effect of different thresholds applied to the pulses with relative ease, without having to account for the pulse amplitude.
The thresholds applied to the pulses were optimised for each measurement condition, choosing the combination that resulted in the best time resolution.
The time resolution is defined as the standard deviation of the distribution of the time difference ($\Delta t$) between the pulses.
The crossing time of the thresholds was determined using a linear interpolation between two points of the waveform.
Since the difference in the amplitude of the pulses measured in the same waveform was less than 5\%, the time jitter of the sensor and readout electronics for a single pulse (or single pulse time resolution) can be calculated by dividing the two pulses time resolution by $\sqrt{2}$.
The single pulse time resolution of the detectors is shown in figure\,\ref{fig:timeRes2x2} as a function of bias and fluence.
The jitter of the sensor irradiated to $\Phi_{eq} = 10^{15}$\,cm$^{-2}$ lies between 508 and 553\,ps.
These values are not shown in the figure for clarity.
The other sensors show a trend of decreasing jitter as a function of bias voltage.
The behaviour is however not monotonous.

%1e15 -> time res 508-553 ps

\begin{figure}
  \centering
  \includegraphics[width = 0.6 \columnwidth]{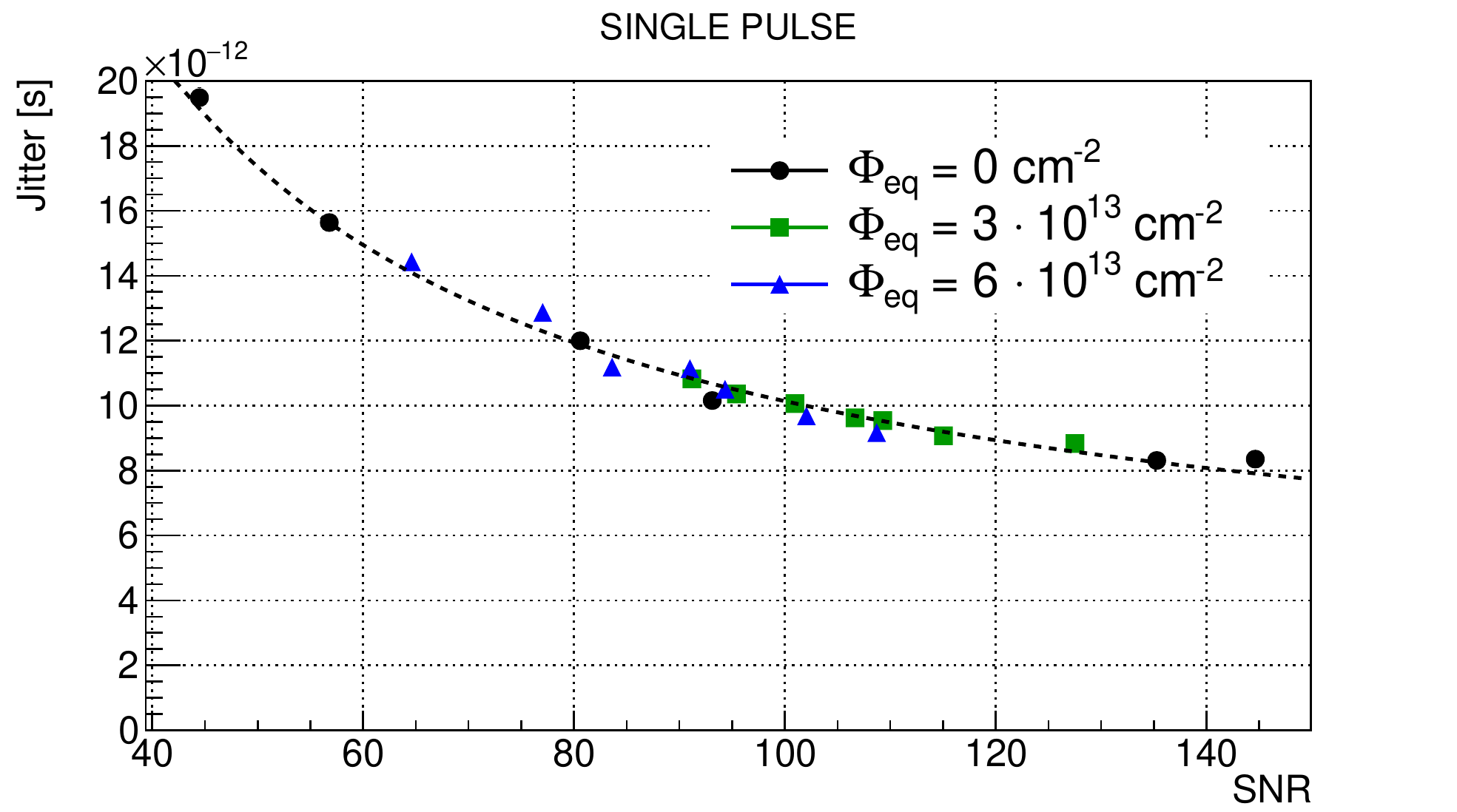}
  \caption{Single pulse time resolution of the $2 \times 2$~mm$^2$ APDs as a function of signal to noise ratio and fluence measured at $-20^\circ$C. The laser intensity corresponds to 0.8\,MIPs. An amplification of 40\,dB was used.}
  \label{fig:timeRes2x2_snr}
\end{figure}

The jitter is found to scale approximately as 1/SNR, as expected from the relation: $\textrm{jitter} \propto \textrm{rise time} / \textrm{SNR}$.
Figure\,\ref{fig:timeRes2x2_snr} shows the jitter as a function of SNR.
The dashed line represents a 1/SNR behaviour.

The jitter of the sensors is not degraded by the neutron-induced radiation damage corresponding to a fluence of at least $\Phi_{eq} = 6 \cdot 10^{13}$\,cm$^{-2}$.
These results are expected to hold for timing measurements of charged particles.
The overall resolution is however expected to worsen due to the fluctuations of the amount of charge deposited per unit length along the particle path.
These fluctuations, also known as {\em Landau noise}\,\cite{cartiglia2017}, can influence the leading edge of the sensor's signal and therefore worsen the time resolution.

\section{Uniformity Study of $8 \times 8$~mm$^2$ APDs using an Infrared Laser}
\label{sec:unif8x8laser}

\begin{figure*}[h]
  \centering
  \subfloat[Amplitude]{\includegraphics[width = 0.49 \textwidth]{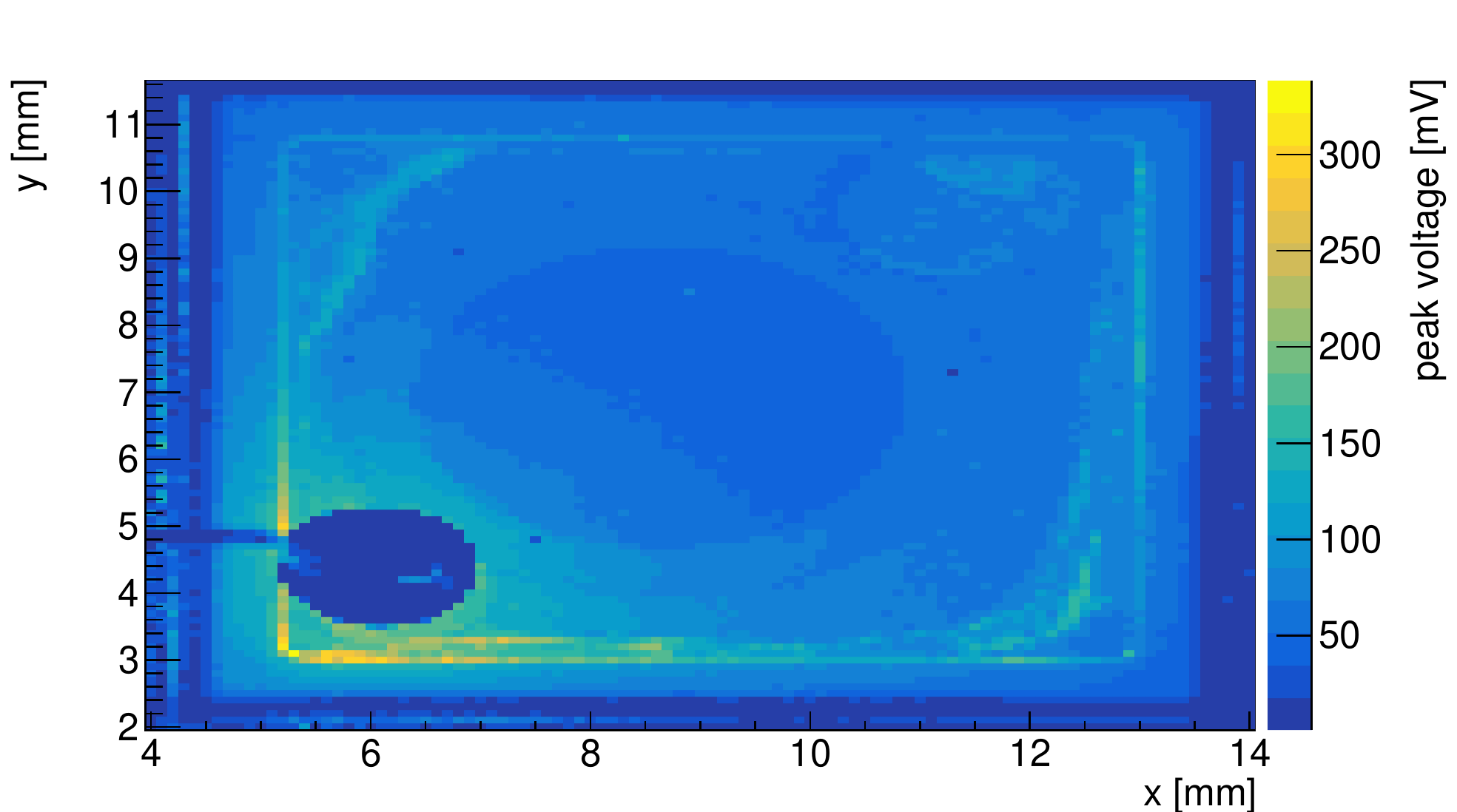}}\hfill
  \subfloat[Integrated signal]{\includegraphics[width = 0.49 \textwidth]{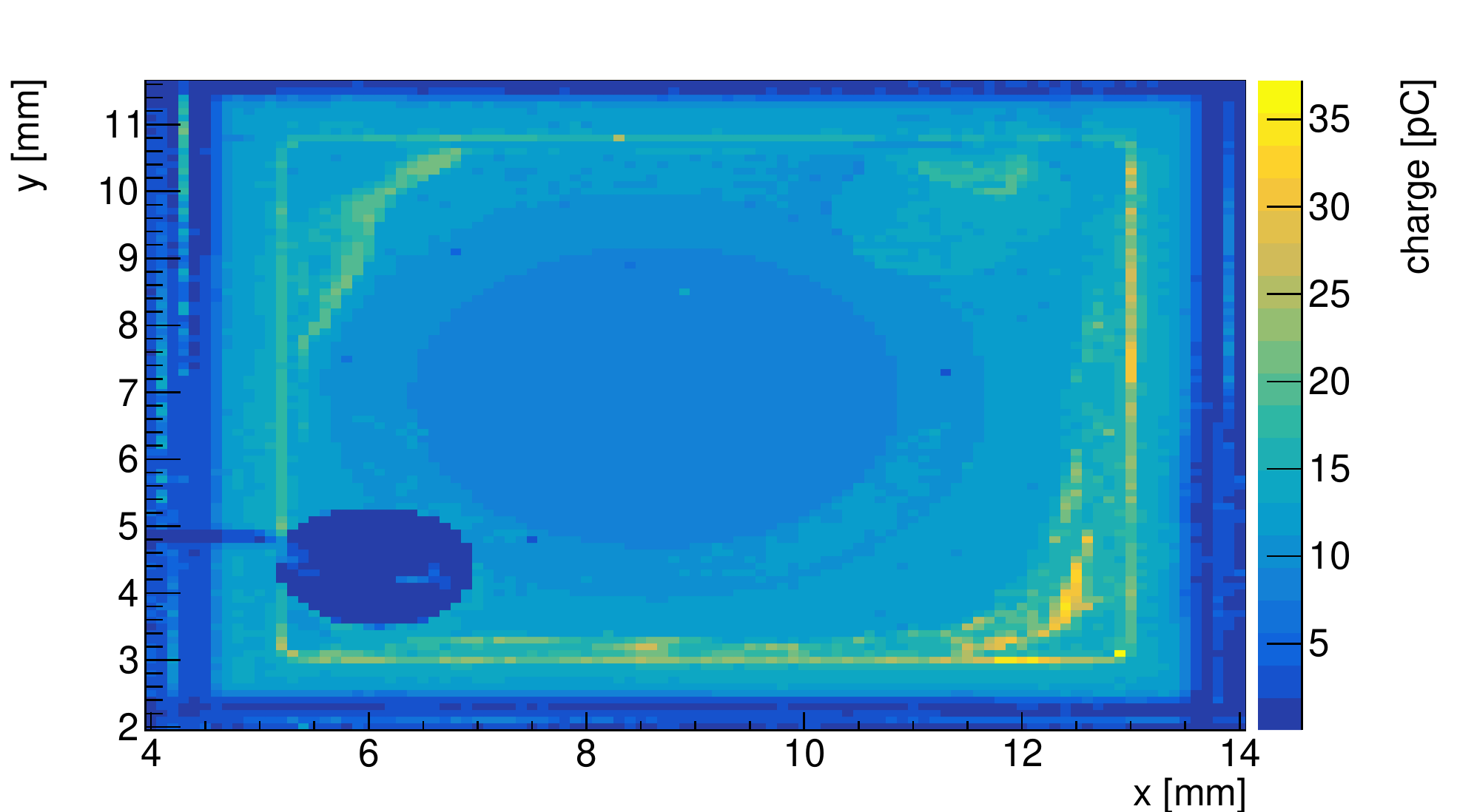}}\\
  \subfloat[Ratio]{\includegraphics[width = 0.49 \textwidth]{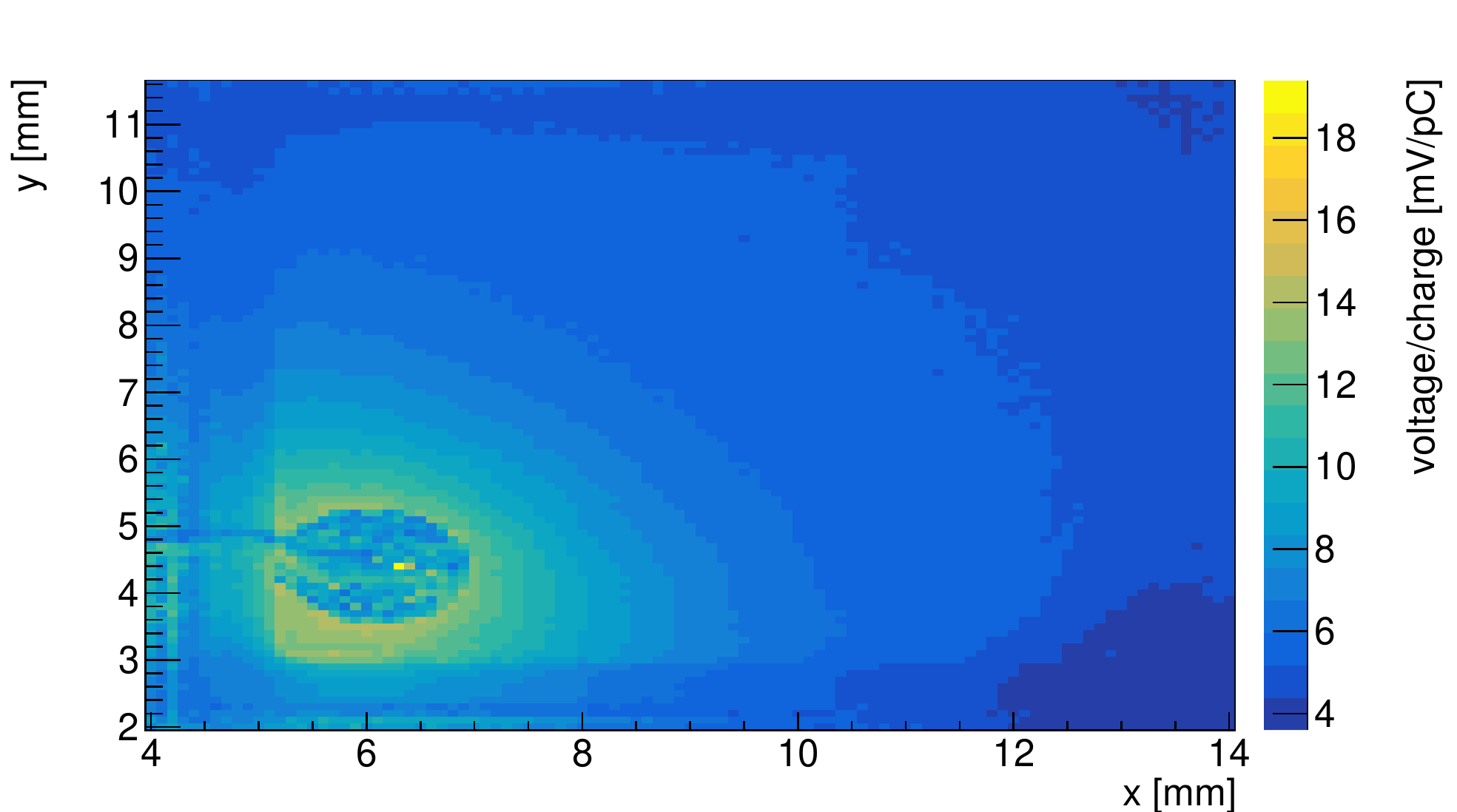}}\hfill
  \subfloat[Ratio vs distance]{\includegraphics[width = 0.49 \textwidth]{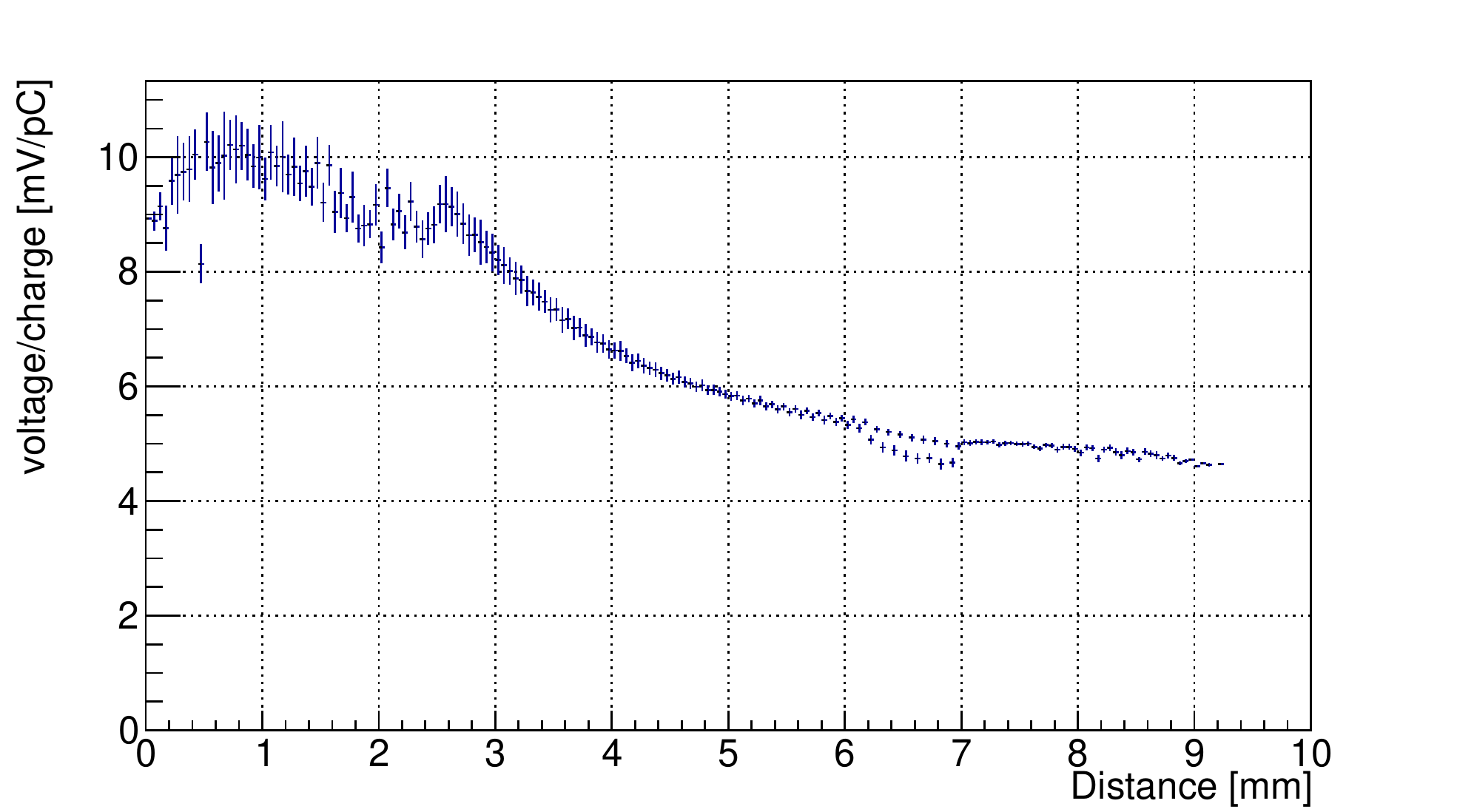}}\\
  \caption{Signal amplitude {\bf (a)}, integral {\bf (b)}, and ratio between amplitude and integral {\bf (c)} of a $8 \times 8$~mm$^2$ APD as a function of the $x-y$ impact point of the laser beam on the sensor. {\bf (d)} Ratio between amplitude and integral as a function of distance from the electrical contact of the sensor. The sensor was biased at 1700\,V, at a temperature of $-20^\circ$C. The laser intensity corresponds to 15 MIPs. An amplification of 10\,dB was used (colour online).}
  \label{fig:8x8unif_noMetal}
\end{figure*}

The uniformity of response of the $8 \times 8$~mm$^2$ APDs was studied using an infrared laser.
The amplitude and integrated signal are shown in figure\,\ref{fig:8x8unif_noMetal} as a function of the position where the laser illuminated the detector.
The laser spot was moved across the sensor in 100\,$\mu$m steps.
The sensor was mounted with the n-side facing the PCB, and the laser shone on the p-side.
The sensor was electrically connected using opaque conductive paint.
The point of contact on the p-side can be seen at the coordinate (6, 4.5)\,mm.
Since the absorption length of the light used in the measurements is greater than the sensor thickness, several features of the surface below the sensor can be seen in the figure.
The square mesa structure and the conductive paint used to glue the detector to the PCB used for the measurements can be seen as areas of increased amplitude and signal integral due to their reflection of the light toward the sensor.
The PCB has a hole below the detector centre, resulting in decreased values of amplitude and signal integral.
Since the signal amplitude is affected by both the deposited charge in the detector and the detector properties, the amplitude was normalised using the integrated signal in order to mitigate the former effect.
The ratio between the amplitude and the integrated signal is shown in figure\,\ref{fig:8x8unif_noMetal}.
The effects of the surface below the sensor are reduced and a dependency of the ratio between amplitude and charge on the distance between the laser spot and the point where the signal is collected from the sensor is shown.
The amplitude of the signal decreases with increasing distance from the electrical contact.
This effect is attributed to the non-negligible resistivity of the conductive layers applied to the detector.
The dependency of the signal amplitude on the detector position can result in a worsened time resolution.

The two approaches taken to reduce this source of non-uniformity are described in section\,\ref{sec:samples}.
%% The results obtained with the AC-coupled mesh readout are shown in section\,\ref{sec:tb8x8}.
The DC-coupled metallisation resulted in a spread of the ratio between amplitude and charge of less than 2\% over a distance of 7\,mm between the points illuminated by the laser.
No dependency on the distance between illumination and electrical contact was observed.
This measurement was performed on a sensor biased to 1800\,V at a temperature of $20^\circ$C.
The metallisation was performed at the clean room facility of CMi-EPFL\,\cite{cmi}.
The structure consists of an aluminium grid on the n-side and a continuous aluminium layer with an opening at the detector centre on the p-side of the detector.
This configuration allows the illumination of the detector centre without reflections.

\section{Timing Performance of $8 \times 8$~mm$^2$ APDs using an Infrared Laser}
\label{sec:timing8x8laser}

\begin{figure}
  \centering
  \includegraphics[width = 0.6 \columnwidth]{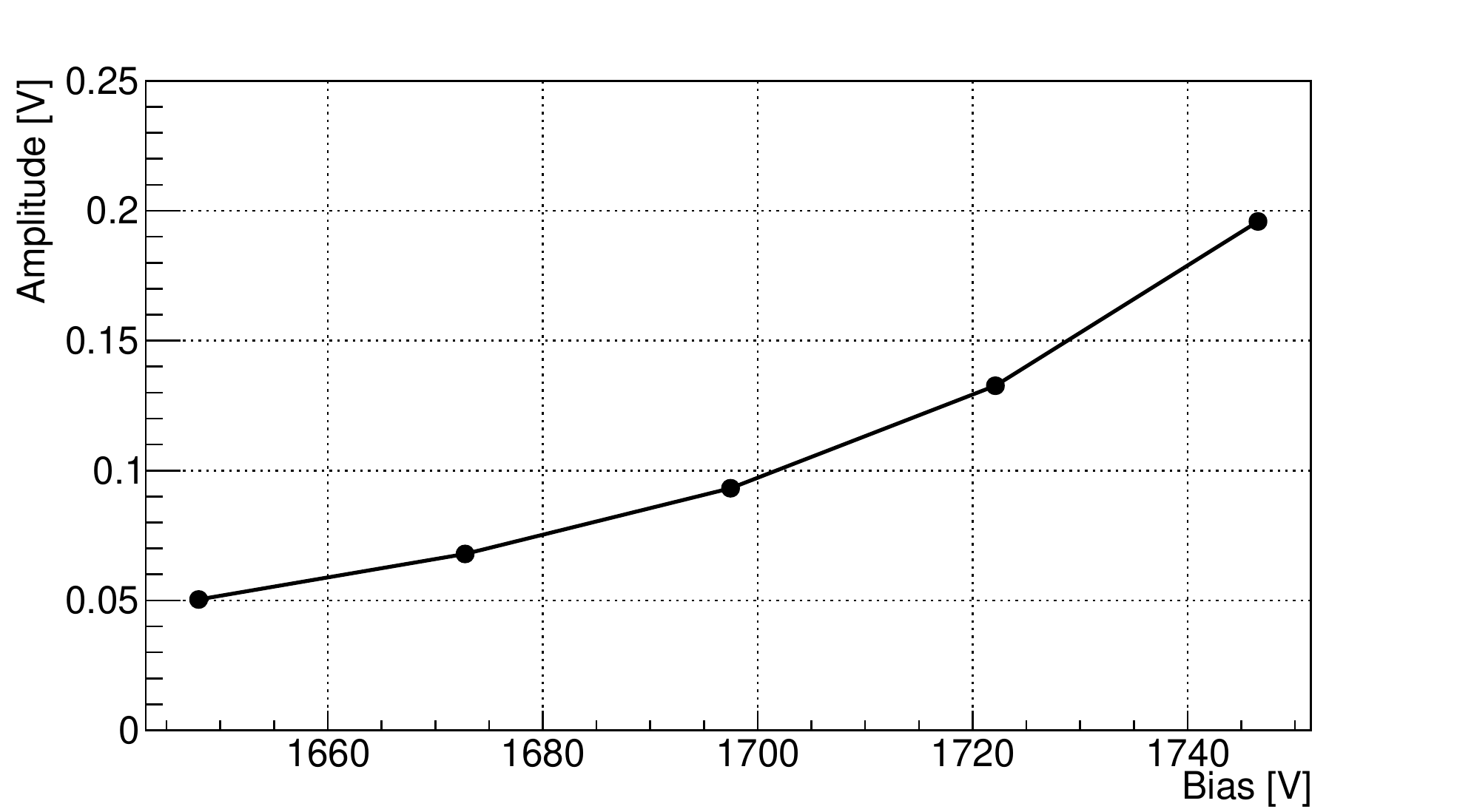}
  \caption{Average amplitude of the metallised $8 \times 8$~mm$^2$ APD signal as a function of bias voltage measured at $20^\circ$C. The laser intensity corresponds to 0.8\,MIPs. An amplification of 40\,dB was used.}
  \label{fig:ampli8x8metal}
\end{figure}

\begin{figure}
  \centering
  \includegraphics[width = 0.6 \columnwidth]{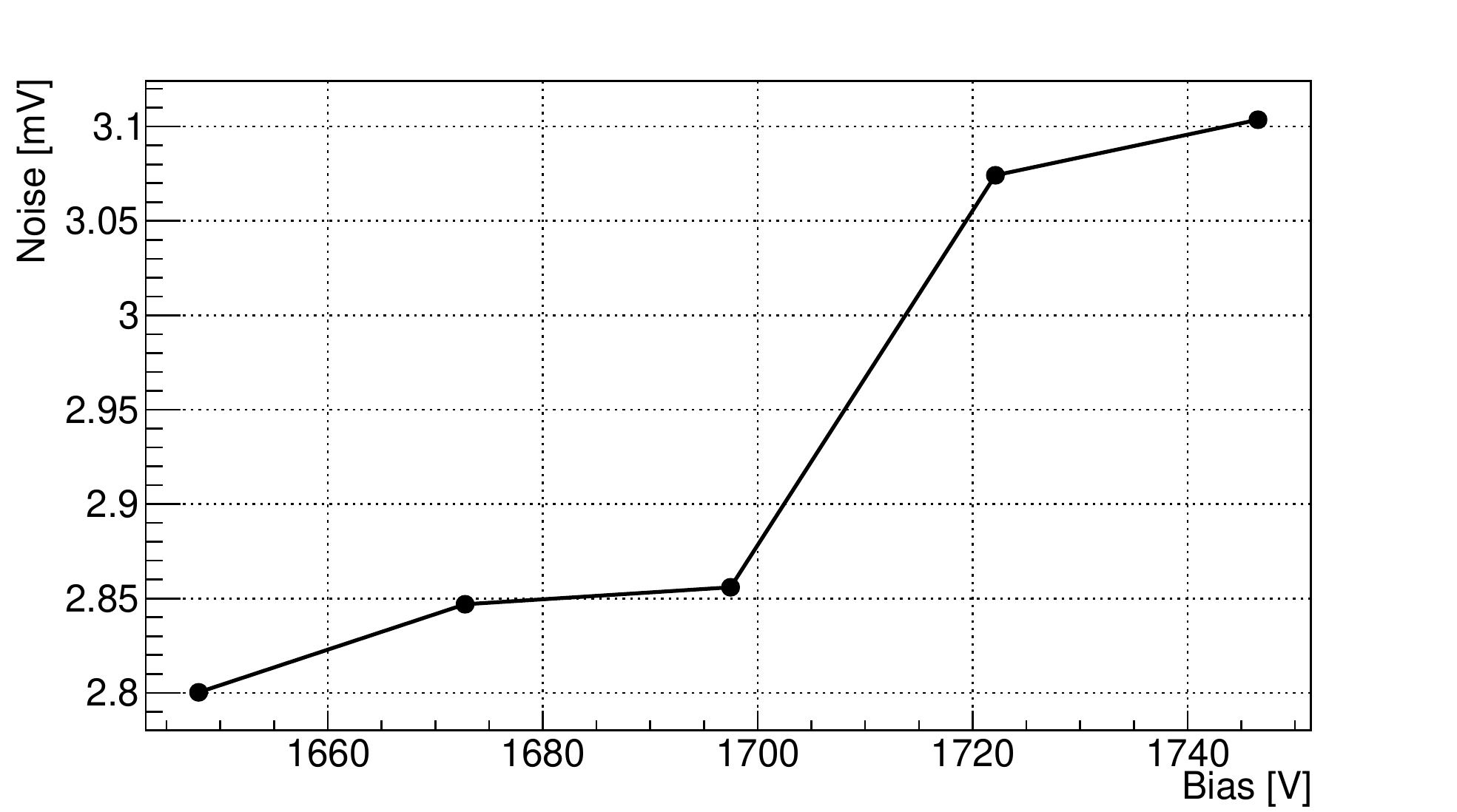}
  \caption{$8 \times 8$~mm$^2$ APD's noise as a function of bias voltage measured at $20^\circ$C. An amplification of 40\,dB was used.}
  \label{fig:noise8x8metal}
\end{figure}

The time resolution of the metallised $8 \times 8$~mm$^2$ APDs with DC-coupled readout was studied using the same setup and procedure used for the characterisation of the irradiated $2 \times 2$~mm$^2$ APDs in section\,\ref{sec:irrad2x2}.
A light intensity corresponding to 0.8\,MIPs and an amplification of 40\,dB were used.
The light was shone in the centre of the detector.
The signal amplitude and noise are shown in figure\,\ref{fig:ampli8x8metal} and\,\ref{fig:noise8x8metal}, respectively.
The signal amplitude is smaller than the one of the $2 \times 2$~mm$^2$ devices shown in section\,\ref{sec:irrad2x2}.
This is a consequence of the lower gain of the device under test and its larger capacitance.
The difference in temperature between the measurements presented in this section and the ones of section\,\ref{sec:irrad2x2} does not explain the observed difference in gain.
%The noise is also smaller than the one of the $2 \times 2$~mm$^2$ devices.
Both signal amplitude and noise increase as a function of bias voltage.
The signal to noise ratio is lower than the one of the non-irradiated $2 \times 2$~mm$^2$ APD presented in section\,\ref{sec:irrad2x2}, ranging from 18 to 62.

\begin{figure}
  \centering
  \includegraphics[width = 0.6 \columnwidth]{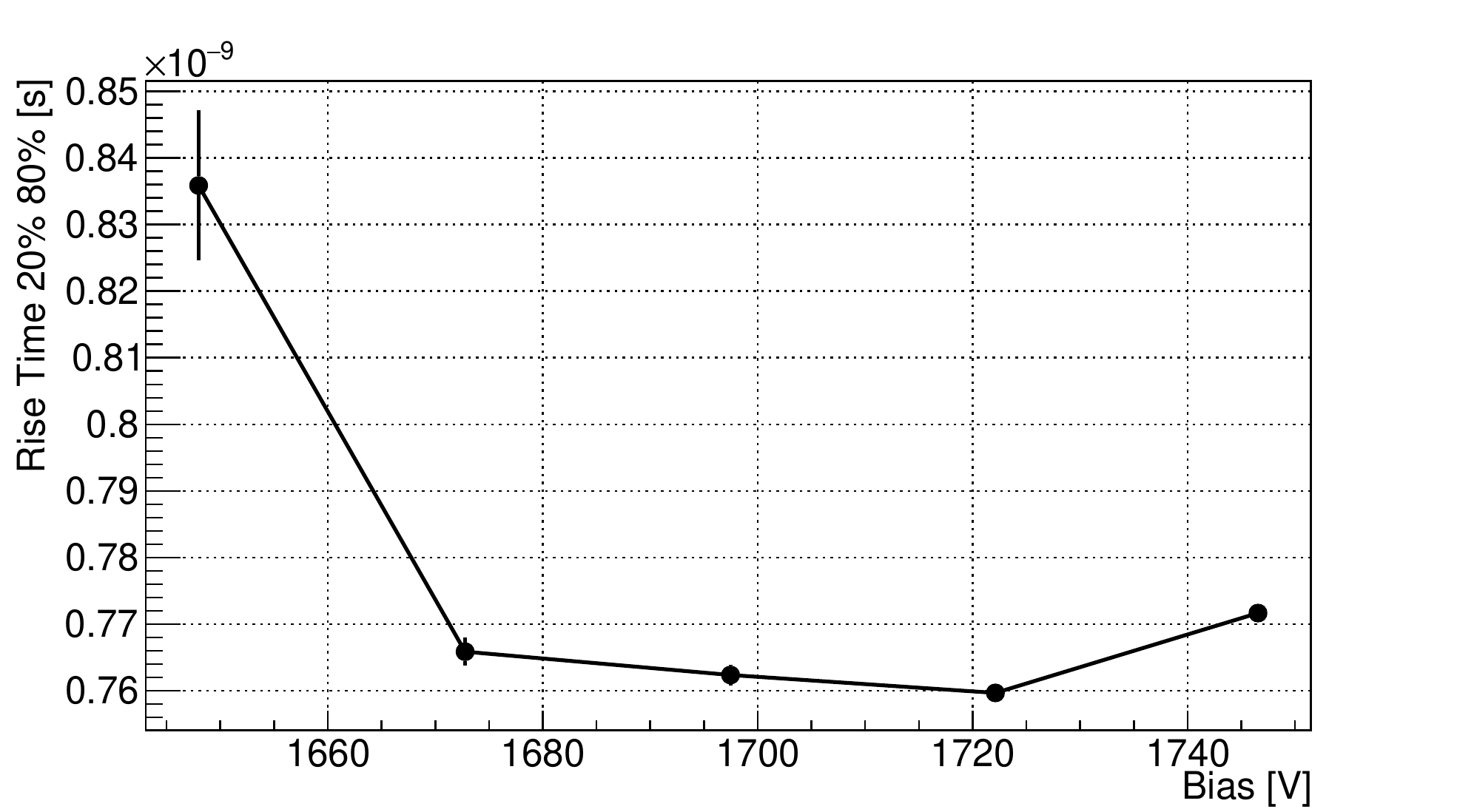}
  \caption{20\%-to-80\% rise time of the metallised $8 \times 8$~mm$^2$ APD signal as a function of bias voltage measured at $20^\circ$C. The laser intensity corresponds to 0.8\,MIPs. An amplification of 40\,dB was used.}
  \label{fig:riseTime8x8metal}
\end{figure}

The 20\%-to-80\% rise time of the signal is shown in figure\,\ref{fig:riseTime8x8metal}.
The values are bigger than the ones of the $2 \times 2$~mm$^2$ devices.
This is a consequence of the bigger sensor capacitance.

The jitter was calculated with the same procedure presented in section\,\ref{sec:irrad2x2}.
The jitter decreases with increasing bias voltage and is found to scale with a 1/SNR behaviour.

\begin{figure}
  \centering
  \includegraphics[width = 0.6 \columnwidth]{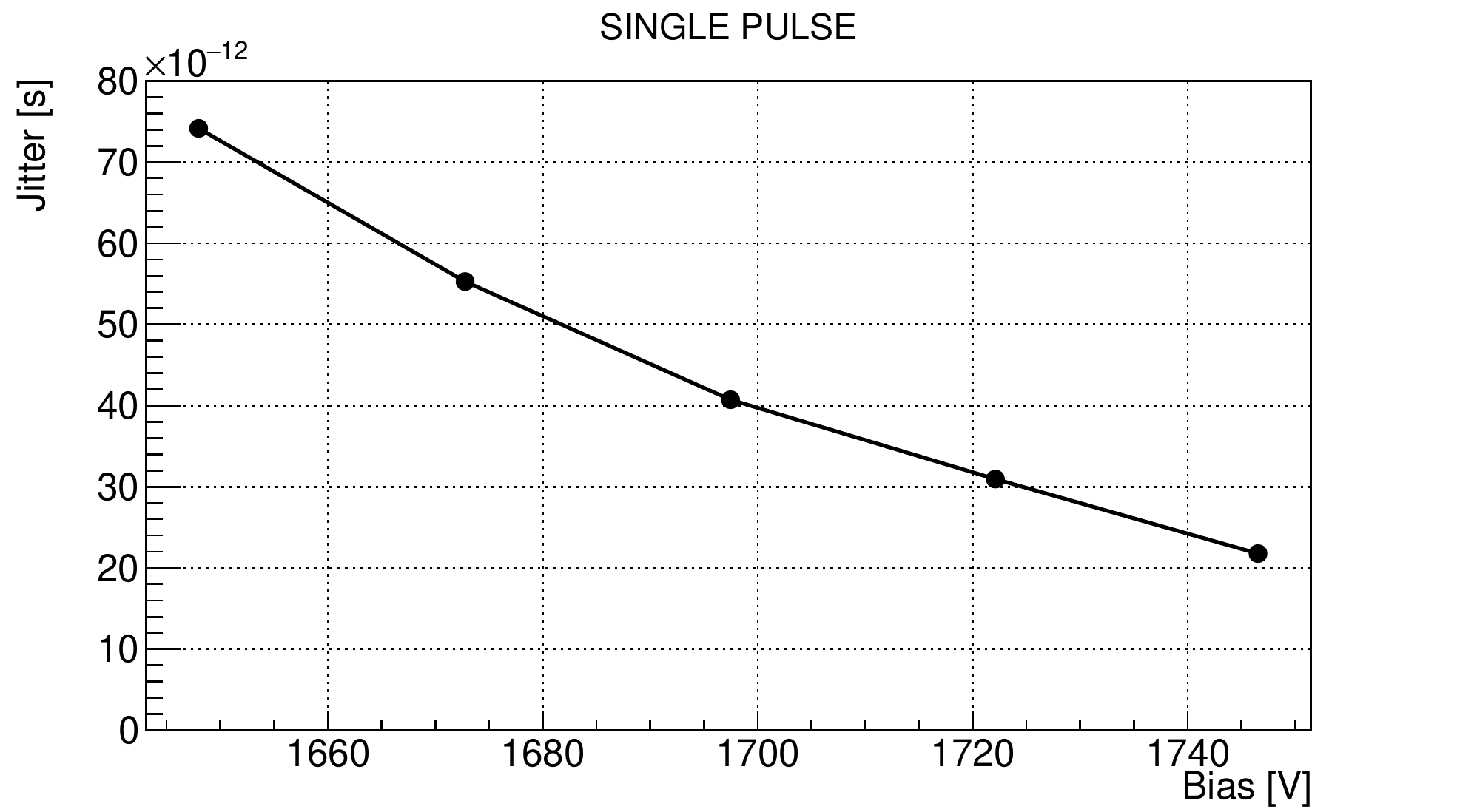}
  \caption{Single pulse time resolution of the metallised $8 \times 8$~mm$^2$ APDs as a function of bias voltage measured at $20^\circ$C. The laser intensity corresponds to 0.8\,MIPs. An amplification of 40\,dB was used.}
  \label{fig:timeRes8x8metal}
\end{figure}

The metallised $8 \times 8$~mm$^2$ APDs often did not allow for a stable operation when biased above 1700\,V.
The source of the instability is under investigation.
This instability influenced the choice of values of bias voltage used in this section.

\section{Beam Tests of $8 \times 8$~mm$^2$ APDs with AC-coupled Readout}
\label{sec:tb8x8}

The performance of AC-coupled mesh APDs was characterised in two beam tests.
One of the beam tests focused on the measurement of the uniformity of response of the APD, enabled by the presence of a beam telescope.
The focus of the second beam test was the measurement of the AC-coupled mesh APDs' time resolution, enabled by a time reference detector (MCP-PMT\footnote{Micro Channel Plate Photomultiplier.}) covering the full area of the APD.

\paragraph{Uniformity of response}
AC-coupled mesh readout $8 \times 8$~mm$^2$ APDs were characterised in terms of their uniformity of response using a 150\,GeV muon beam in the H4 beam line at the CERN SPS\,\cite{h4page}.
For these measurements, a fast transimpedance amplifier with $\approx 10$\,$\Omega$ effective input impedance was used\,\cite{whiteACES2014}.
The amplifier used for the uniformity measurements was realised using discrete components, and its bandwidth was 1\,GHz.
The APD and amplifier were operated at room temperature ($\approx20^\circ$C) during the measurements, the APD bias voltage was 1750\,V.
A 2.5\,GHz 20\,GSa/s oscilloscope was used to digitise the waveforms.
The measurements were carried out within the infrastructure of PICOSEC\,\cite{bortfeld2018}.
In particular, the PICOSEC setup provided particle tracking with a resolution of 40\,$\mu$m.
The reference frame used in the beam test was right handed, with the beam parallel to the $z$ axis, and the particles moving from positive to negative $z$ values.
The $x$ and $y$ axes were horizontal and vertical, respectively, their origin corresponds with the centre of the APD.

\begin{figure*}[h]
  \centering
  \subfloat[Amplitude distribution]{\includegraphics[width = 0.49 \textwidth]{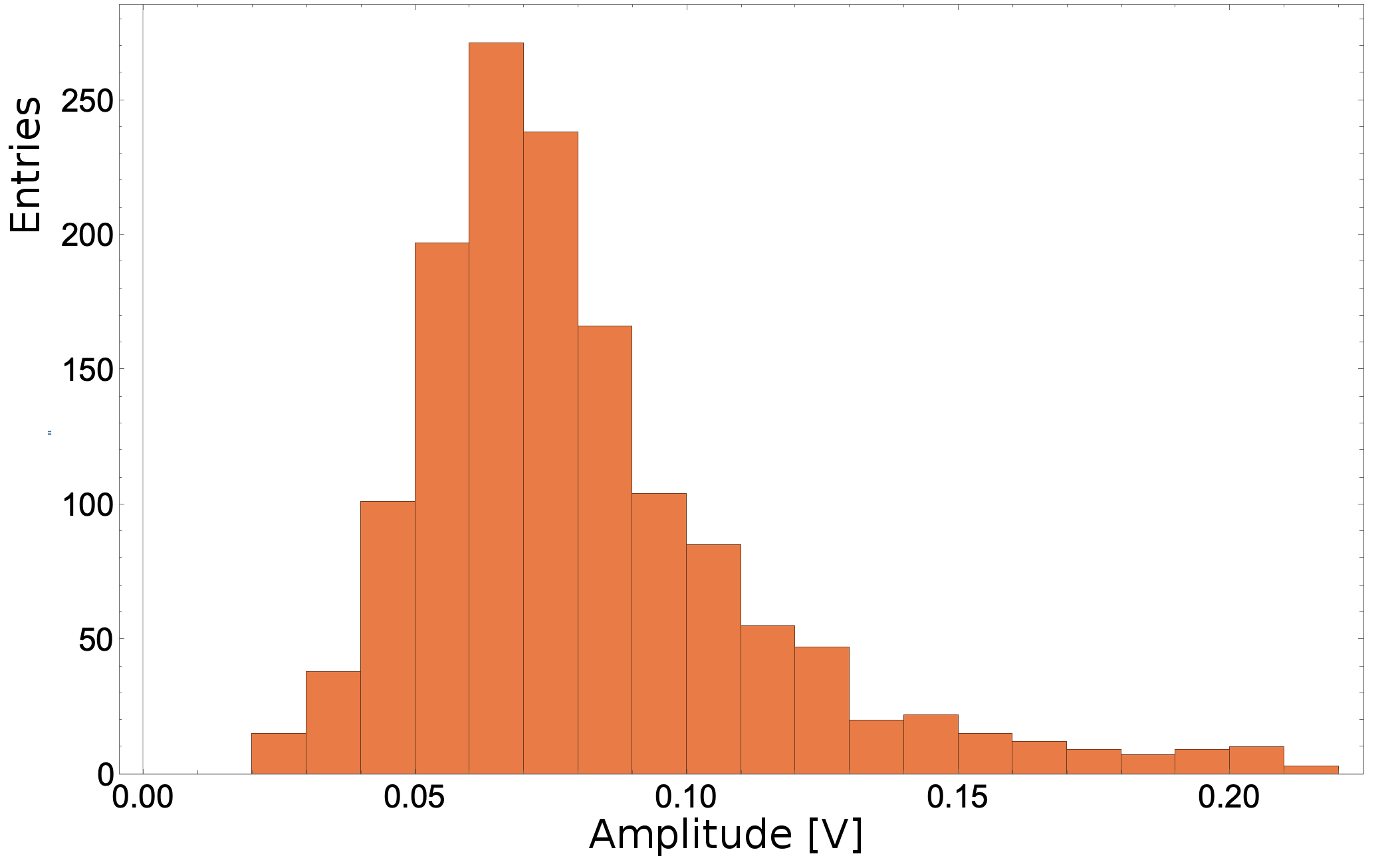}\label{fig:ampli8x8distrTB}}\hfill
  \subfloat[Rise time definition]{\includegraphics[width = 0.49 \textwidth]{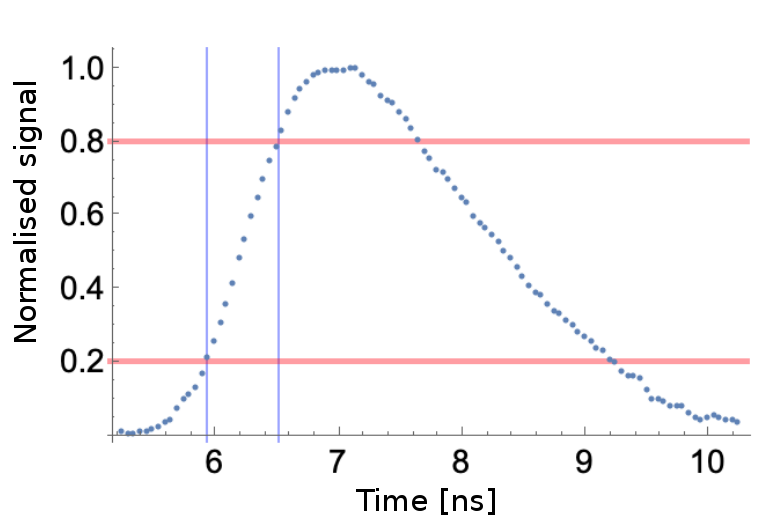}\label{fig:riseTime8x8TB}}\\
  \subfloat[Mean amplitude]{\includegraphics[width = 0.49 \textwidth]{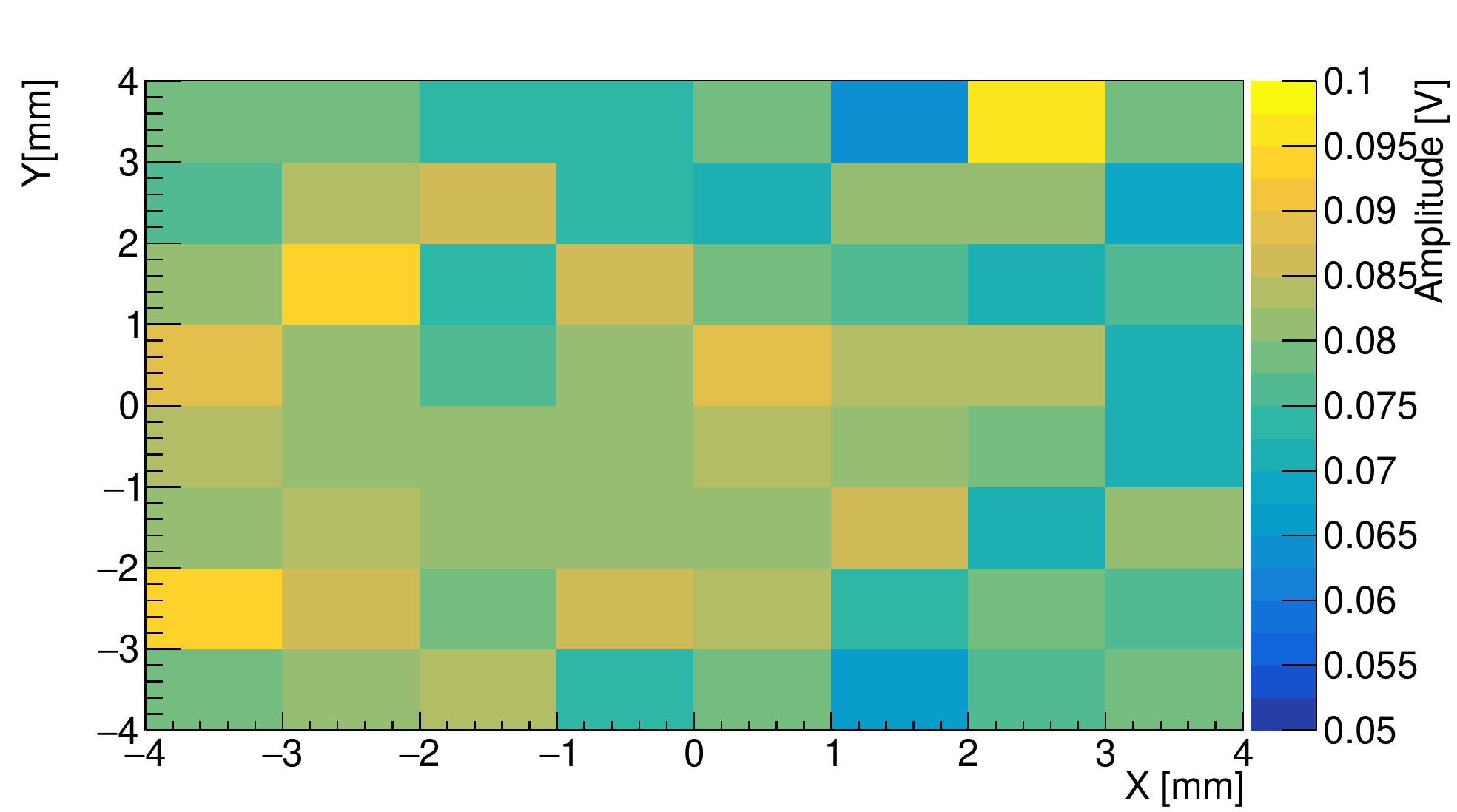}\label{fig:ampli8x8mapTB}}\hfill
  \subfloat[Mean rise time]{\includegraphics[width = 0.49 \textwidth]{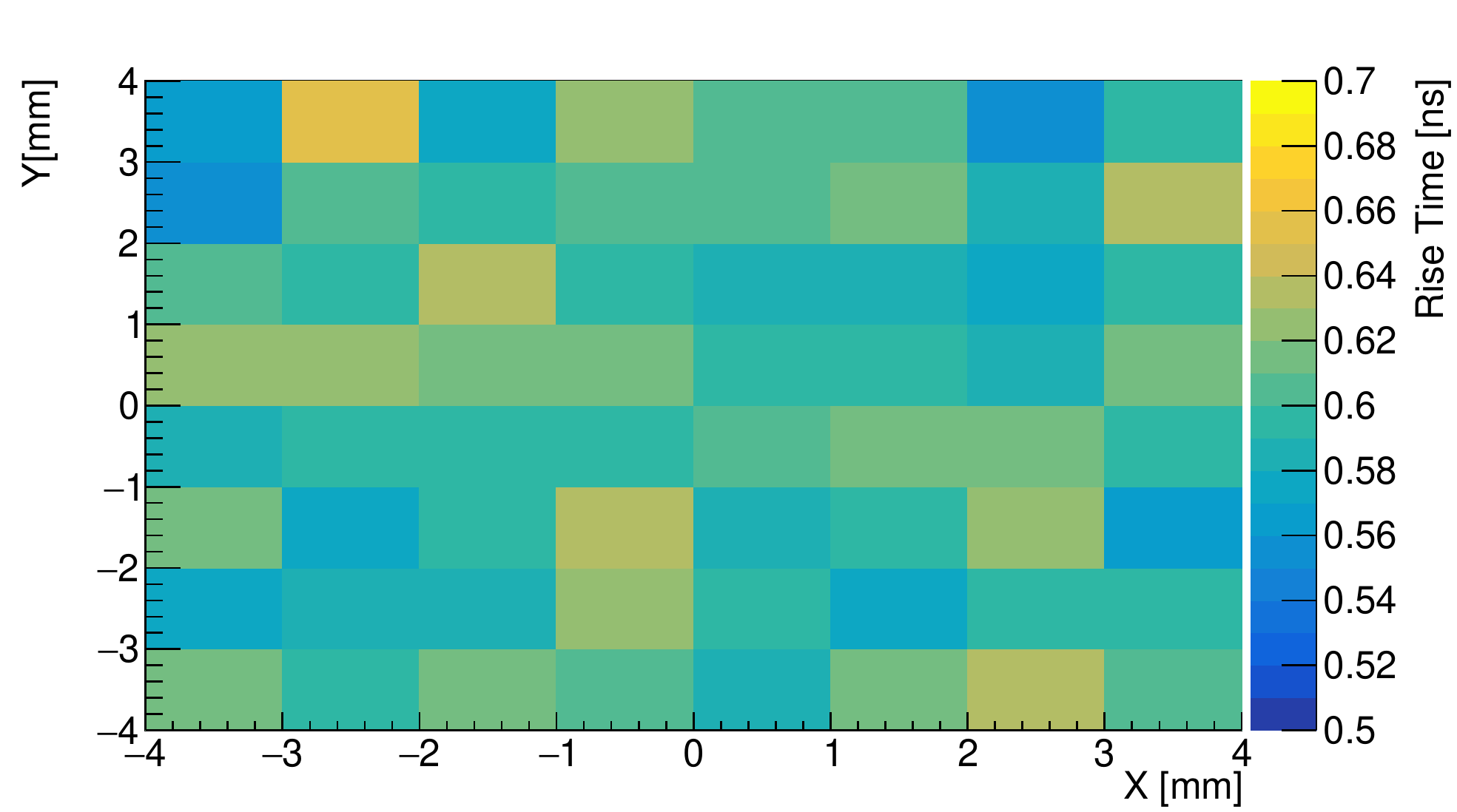}\label{fig:riseTime8x8mapTB}}
  \caption{Amplitude distribution \textbf{(a)} of the AC-coupled mesh readout APD at the beam test. Normalised signal of a typical event and definition of the 20-to-80\% rise time \textbf{(b)}. Mean amplitude \textbf{(c)} and rise time \textbf{(d)} as a function of the particles impact position. The APD was biased to 1750\,V and kept close to room temperature. The measurements were performed using 150\,GeV muons (colour online).}
  \label{fig:uniformityRiseTimeAmpliTB}
\end{figure*}

The results of the uniformity study are shown in figure\,\ref{fig:uniformityRiseTimeAmpliTB}.
Figure\,\ref{fig:ampli8x8distrTB} shows the amplitude distribution of the APD signal for the data used in the uniformity study.
The most probable value (MIP peak) is 65\,mV at the oscilloscope input, for the 1750\,V bias applied to the sensor.
The average amplitude as a function of the particles impact position on the APD is shown in figure\,\ref{fig:ampli8x8mapTB}.
The amplitude is uniform over the detector's active area.
The normalised signal of a typical event is shown in figure\,\ref{fig:riseTime8x8TB}.
Figure\,\ref{fig:riseTime8x8TB} also depicts the definition of 20-to-80\% rise time used in this section.
The average rise time as a function of the particles impact position is shown in figure\,\ref{fig:riseTime8x8mapTB}.
The values of each bin shown in figure\,\ref{fig:riseTime8x8mapTB} lie within $\pm8.5$\% of the overall average rise time of 600\,ps.
The rise time is uniform over the detector's active area, and its values do not show any particular trend.

\begin{figure}
  \centering
  \includegraphics[width = 0.6 \columnwidth]{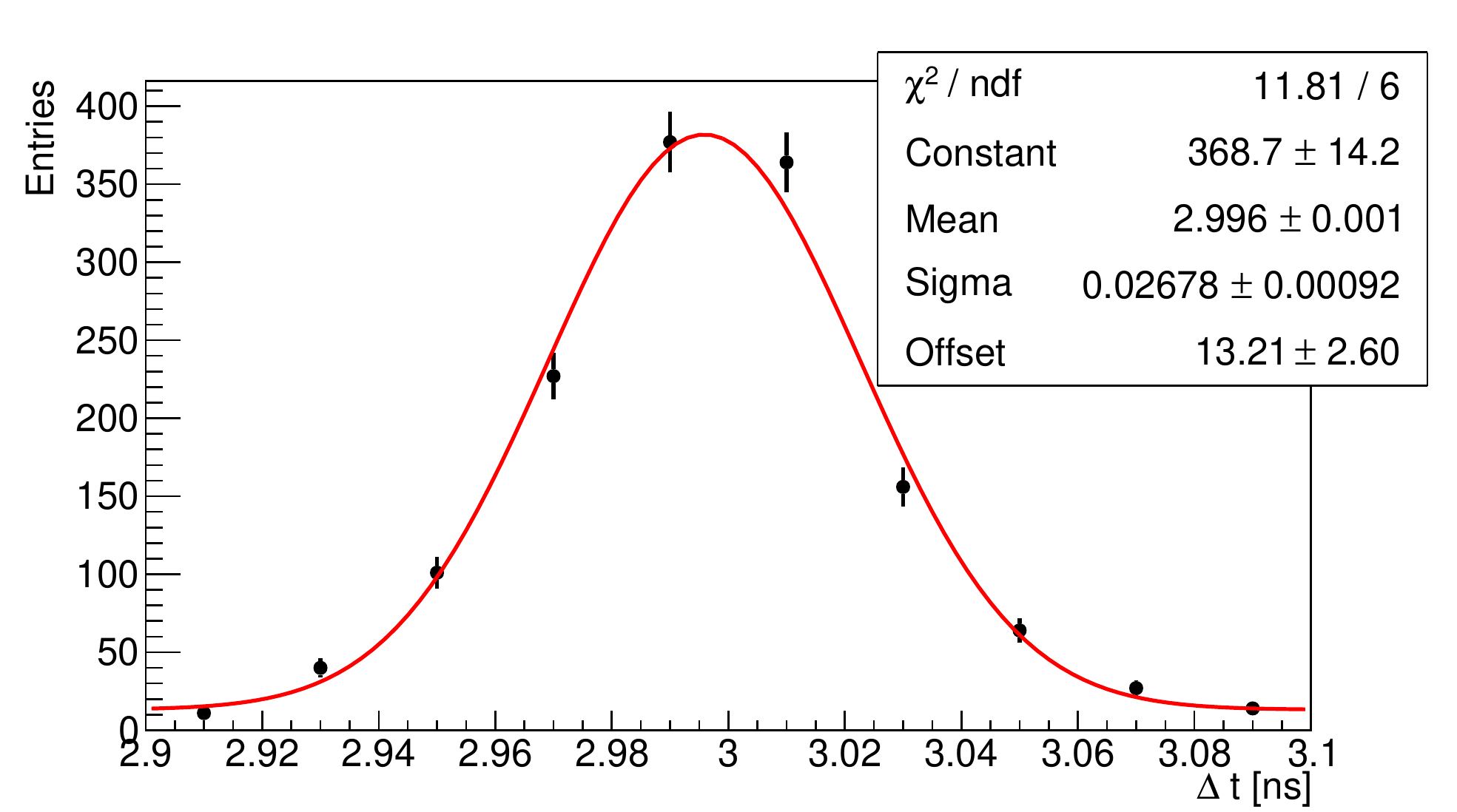}
  \caption{Distribution of the time difference between the MCP-PMT trigger signal and the mesh APD. The distribution is described using the sum of a Gauss function with a constant offset. The measurement was performed using 120\,GeV protons. The APD was operated at 1800\,V.}
  \label{fig:dtDistrTB}
\end{figure}

\paragraph{Time resolution}
The time resolution measurements for AC-coupled mesh readout $8 \times 8$~mm$^2$ APDs were performed using a 120\,GeV proton  beam at the Fermilab test beam facility\,\cite{ftbfPage}.
The proton beam illuminated the whole active area of the APD, that was operated close to room temperature, at a bias of 1800\,V.
The setup consisted of an MCP-PMT\footnote{The model used was Hamamatsu R3809U-50.} used as trigger and time reference, and the APD.
The particles are detected by the MCP-PMT as they emit Cherenkov light in its quartz window.
The signal from the mesh APD was amplified by an ASIC (FASTAMP) fabricated in a sub micron Silicon Germanium bipolar process designed by the University of Pennsylvania Instrumentation group\,\cite{whiteACES2014}.
Each channel consists of a pseudo-differential preamplifier followed by a short time constant shaping stage with two selectable gains of 2 and 5\,mV/fC and a differential driver output stage.
The results reported here are from the 5\,mV/fC gain setting.
The pseudo-differential input stage and sub-ns peaking time results in an equivalent input noise charge of $\approx$ 3000\,e$^-$ for a sensor capacitance of $\approx$ 20\,pF.    
The signals were digitised using a 2.5\,GHz 10\,GSa/s oscilloscope triggered using the MCP-PMT signal.
As the leading edge of the MCP-PMT signal contained too few points for being directly used as a time reference, the measurement of time difference ($\Delta t$) is done with respect to the oscilloscope trigger.
The time walk effect due to the varying amplitude of the MCP-PMT signal was corrected in the offline analysis.
The mesh APD signal used in the analysis is obtained by combining the digitised differential outputs of the ASIC amplifier.
As a phase difference of about 50\,ps is present between the differential signals, their combination effectively improved the sampling of the APD signal to 20\,GSa/s.
The events with an amplitude larger than 40\,mV were used in the analysis, while the MIP peak was 55\,mV.
To reduce the influence of the noise on the APD signal, a 500\,MHz digital low-pass filter was applied to the signal.
A power law was then fitted to the signal's leading edge.
The fit function, together with the amplitude information, were then put as input to a CFD algorithm.
The threshold used in the analysis is 0.2.
The distribution of the time difference obtained with this analysis is shown in figure\,\ref{fig:dtDistrTB}.
The distribution is described using a Gauss function summed to a constant offset.
The use of the offset parameter improves the description of the data.
The time resolution of the system mesh APD MCP-PMT is found to be $27 \pm 1$\,ps.

\section{Summary}
\label{sec:summary}

For their operation at HL-LHC, the ATLAS and CMS experiments foresee detector upgrades that add timing information of reconstructed charged particles, aiming at a time resolution of around 30\,ps.
The detectors used in these systems will be exposed to irradiation fluences of up to $\Phi_{eq} = 10^{15}$\,cm$^{-2}$ for the target integrated luminosity of HL-LHC of 3000\,fb$^{-1}$.

The dependence of the gain of deep diffused APDs on bias voltage and temperature was studied using a pulsed blue LED.
A scaling law for the gain was found, allowing to determine which change in bias voltage is necessary to maintain the same gain when varying the APD temperature.

The parameters influencing the time resolution of deep diffused APDs with DC-coupled readout, as well as the uniformity of response of these devices, were measured.

Deep diffused APDs with an active area of $2 \times 2$\,mm$^2$ produced by RMD were irradiated with reactor neutrons up to a fluence of $\Phi_{eq} = 10^{15}$\,cm$^{-2}$.
The uniformity of response was studied using a pulsed infrared laser.
The sensitive area of the detectors is affected by irradiation, maintaining a uniform response at the centre of the sensors.
Current, amplitude, and timing characteristics of the detectors were measured at a temperature of $-20^\circ$C, before and after irradiation.
From the current-voltage characteristic it was found that the bulk current of the detectors increases with irradiation and that the gain decreases with irradiation.
The latter observation is supported by measurements of the APDs signal performed using a pulsed infrared laser illuminating the centre of the detectors.
The time jitter of the APDs was determined using a pulsed infrared laser with an intensity corresponding to 0.8\,MIPs per pulse.
The laser illuminated the centre of the detectors.
The jitter is not degraded by exposure to fluences of up to at least $\Phi_{eq} = 6 \cdot 10^{13}$\,cm$^{-2}$.
The jitter of the device irradiated to $\Phi_{eq} = 3 \cdot 10^{14}$\,cm$^{-2}$ could not be determined due to its unstable behaviour, while the device irradiated to $\Phi_{eq} = 10^{15}$\,cm$^{-2}$ did not show any gain.

The uniformity of response of deep diffused APDs with an active area of $8 \times 8$\,mm$^2$ was studied using an infrared laser.
The amplitude of the signal shows a dependency on the distance between the point where the laser is shone and the electrical contact of the sensor.
The deposition of an aluminium layer on the detector surfaces improved the uniformity.
The timing characteristics of the metallised $8 \times 8$\,mm$^2$ APD were studied using a pulsed infrared laser with an intensity corresponding to 0.8\,MIPs per pulse.
The time jitter was found to be 22\,ps at a voltage of 1750\,V.
The metallised sensors often showed instabilities when biased above 1700\,V.
The origin of these instabilities is under investigation.

The AC-coupled mesh readout of $8 \times 8$\,mm$^2$ APDs resulted in a satisfactory uniformity of response as studied using a high energy muon beam.
This readout relies on the mesh providing a low impedance path for the signal induced by the charge carriers motion within the silicon.
The rise time and signal amplitude do not show a significant dependence on the impact position of the particles on the sensor.
The mesh readout also reduces the capacitance seen by the amplifier, enabling development of fast low noise signal processing front end, which will be discussed elsewhere.
The time resolution of the mesh APD was determined in a dedicated beam test.
Using a proton beam that fully illuminated the active area of the detector, a time resolution of $27 \pm 1$\,ps was measured for a 1800\,V bias.

Deep diffused APDs were characterised for response uniformity and timing performance in laboratory and test beam studies, before and after irradiation.
An extensive set of results is presented.
Further investigation is needed to fully characterise these detectors for their use in the future upgrades of the LHC experiments.

\section*{Acknowledgements}

The work summarised in this paper has been performed within the framework of the RD50 Collaboration.
This project has received funding from the European Union’s Horizon 2020 Research and Innovation programme under Grant Agreement no.\ 654168.
The authors wish to thank J.~Bronuzzi for the help during the clean room operations and in the development of the recipe for the metallisation of the devices.
The authors are grateful to the members of the CERN EP-DT-DD SSD team that developed and maintain the setups used in this work.
S.\ White acknowledges partial support through the US CMS program under DOE contract No.\ DE-AC02-07CH11359.

\bibliography{bibliography}

\begin{thebibliography}{10}
\expandafter\ifx\csname url\endcsname\relax
  \def\url#1{\texttt{#1}}\fi
\expandafter\ifx\csname urlprefix\endcsname\relax\def\urlprefix{URL }\fi
\expandafter\ifx\csname href\endcsname\relax
  \def\href#1#2{#2} \def\path#1{#1}\fi

\bibitem{hlLhcTecDesRep}
{HiLumi LHC Collaboration}, {HiLumi LHC Technical Design Report},
  \url{http://hilumilhc.web.cern.ch/science/deliverables} (2015).

\bibitem{atlasPileup}
{ATLAS Collaboration}, {ATLAS Experiment Public Results},
  \url{https://twiki.cern.ch/twiki/bin/view/AtlasPublic/LuminosityPublicResultsRun2#2018_pp_Collisions}.

\bibitem{cmsPileup}
{CMS Collaboration}, {CMS Luminosity Public Results},
  \url{https://twiki.cern.ch/twiki/bin/view/CMSPublic/LumiPublicResults#2018_proton_proton_collisions}.

\bibitem{cmsMIPtiming}
{CMS Collaboration}, \href{http://cds.cern.ch/record/2296612}{{TECHNICAL
  PROPOSAL FOR A MIP TIMING DETECTOR IN THE CMS EXPERIMENT PHASE 2 UPGRADE}},
  Tech. Rep. CERN-LHCC-2017-027. LHCC-P-009, CERN, Geneva (Dec 2017).
\newline\urlprefix\url{http://cds.cern.ch/record/2296612}

\bibitem{atlasMIPtiming}
{ATLAS Collaboration}, \href{http://cds.cern.ch/record/2623663}{{Technical
  Proposal: A High-Granularity Timing Detector for the ATLAS Phase-II
  Upgrade}}, Tech. Rep. CERN-LHCC-2018-023. LHCC-P-012, CERN, Geneva (Jun
  2018).
\newline\urlprefix\url{http://cds.cern.ch/record/2623663}

\bibitem{rmdAddress}
{Radiation Monitoring Devices Inc.}, 44 Hunt St. Watertown USA,
  \url{http://rmdinc.com/}.

\bibitem{white2014}
S.~White, {R\&D for a Dedicated Fast Timing Layer in the CMS Endcap Upgrade},
  Acta Physica Polonica B Proceedings Supplement 7~(4).

\bibitem{mcclish2004}
M.~McClish, R.~Farrell, F.~Olschner, M.~R. Squillante, G.~Entine, K.~S. Shah,
  Characterization of very large silicon avalanche photodiodes, in: IEEE
  Symposium Conference Record Nuclear Science 2004., Vol.~2, 2004, pp.
  1270--1273 Vol. 2.
\newblock \href {http://dx.doi.org/10.1109/NSSMIC.2004.1462432}
  {\path{doi:10.1109/NSSMIC.2004.1462432}}.

\bibitem{apdPatent}
R.~Farrell, K.~Vanderpuye, \href{http://www.google.com/patents/US7268339}{Large
  area semiconductor detector with internal gain}, {US Patent 7,268,339}
  (Sep.~11 2007).
\newline\urlprefix\url{http://www.google.com/patents/US7268339}

\bibitem{theoryDDAPD}
R.~H. Redus, R.~Farrell, \href{https://doi.org/10.1117/12.245118}{Gain and
  noise in very high-gain avalanche photodiodes: theory and experiment},
  Proc.SPIE 2859 (1996) 288 -- 297.
\newblock \href {http://dx.doi.org/10.1117/12.245118}
  {\path{doi:10.1117/12.245118}}.
\newline\urlprefix\url{https://doi.org/10.1117/12.245118}

\bibitem{jsiIrrad}
L.~Snoj, G.~Žerovnik, A.~Trkov,
  \href{http://www.sciencedirect.com/science/article/pii/S0969804311005963}{{Computational
  analysis of irradiation facilities at the JSI TRIGA reactor}}, Applied
  Radiation and Isotopes 70~(3) (2012) 483 -- 488.
\newblock \href
  {http://dx.doi.org/https://doi.org/10.1016/j.apradiso.2011.11.042}
  {\path{doi:https://doi.org/10.1016/j.apradiso.2011.11.042}}.
\newline\urlprefix\url{http://www.sciencedirect.com/science/article/pii/S0969804311005963}

\bibitem{vlado}
{V. Cindro, Jo\v{z}ef Stefan Institute, Ljubljana}, Private communication.

\bibitem{cividec}
{CIVIDEC Instrumentation GmbH}, Schottengasse 3A Wien Austria,
  \url{http://cividec.at/}.

\bibitem{ugobonoThesis}
S.~Otero~Ugobono, \href{http://cds.cern.ch/record/2649130}{{Characterisation
  and Optimisation of Radiation-Tolerant Silicon Sensors with Intrinsic Gain}},
  presented 21 Nov 2018 (Jul 2018).
\newline\urlprefix\url{http://cds.cern.ch/record/2649130}

\bibitem{kramberger2015}
G.~Kramberger, M.~Baselga, V.~Cindro, P.~Fernandez-Martinez, D.~Flores,
  Z.~Galloway, A.~Gori{\v{s}}ek, V.~Greco, S.~Hidalgo, V.~Fadeyev,
  I.~Mandi{\'{c}}, M.~Miku{\v{z}}, D.~Quirion, G.~Pellegrini, H.-W.
  Sadrozinski, A.~Studen, M.~Zavrtanik,
  \href{https://doi.org/10.1088%2F1748-0221%2F10%2F07%2Fp07006}{Radiation
  effects in low gain avalanche detectors after hadron irradiations}, Journal
  of Instrumentation 10~(07) (2015) P07006--P07006.
\newblock \href {http://dx.doi.org/10.1088/1748-0221/10/07/p07006}
  {\path{doi:10.1088/1748-0221/10/07/p07006}}.
\newline\urlprefix\url{https://doi.org/10.1088%2F1748-0221%2F10%2F07%2Fp07006}

\bibitem{gurimskaya2019}
Y.~Gurimskaya, P.~{Dias de Almeida}, M.~{Fernandez Garcia}, I.~{Mateu Suau},
  M.~Moll, E.~Fretwurst, L.~Makarenko, I.~Pintilie,
  \href{http://www.sciencedirect.com/science/article/pii/S0168900219307181}{Radiation
  damage in p-type epi silicon pad diodes irradiated with protons and
  neutrons}, NIM A\href
  {http://dx.doi.org/https://doi.org/10.1016/j.nima.2019.05.062}
  {\path{doi:https://doi.org/10.1016/j.nima.2019.05.062}}.
\newline\urlprefix\url{http://www.sciencedirect.com/science/article/pii/S0168900219307181}

\bibitem{cartiglia2017}
{N. Cartiglia et al.},
  \href{http://www.sciencedirect.com/science/article/pii/S0168900216304715}{Tracking
  in 4 dimensions}, NIM A 845 (2017) 47 -- 51, proceedings of the Vienna
  Conference on Instrumentation 2016.
\newblock \href {http://dx.doi.org/https://doi.org/10.1016/j.nima.2016.05.078}
  {\path{doi:https://doi.org/10.1016/j.nima.2016.05.078}}.
\newline\urlprefix\url{http://www.sciencedirect.com/science/article/pii/S0168900216304715}

\bibitem{cmi}
{CMi-EPFL}, CH-1015 Lausanne, \url{https://cmi.epfl.ch/}.

\bibitem{h4page}
{H4 Beamline page},
  \url{http://sba.web.cern.ch/sba/BeamsAndAreas/resultbeam.asp?beamline=H4}.

\bibitem{whiteACES2014}
{S. White et. al.}, {Electronics Challenges for HL-LHC pileup Mitigation with
  HyperFast Timing},
  \url{https://indico.cern.ch/event/287628/contributions/1640928/attachments/535335/738096/acesposter_swhite.pdf},
  {ACES 2014 - Fourth ATLAS CMS Electronics Workshop for LHC Upgrades} (2014).

\bibitem{bortfeld2018}
{J. Bortfeldt et. al.},
  \href{http://www.sciencedirect.com/science/article/pii/S0168900218305369}{{PICOSEC:
  Charged particle timing at sub-25 picosecond precision with a Micromegas
  based detector}}, NIM A 903 (2018) 317 -- 325.
\newblock \href {http://dx.doi.org/https://doi.org/10.1016/j.nima.2018.04.033}
  {\path{doi:https://doi.org/10.1016/j.nima.2018.04.033}}.
\newline\urlprefix\url{http://www.sciencedirect.com/science/article/pii/S0168900218305369}

\bibitem{ftbfPage}
{Fermilab test beam facility page}, \url{https://ftbf.fnal.gov/}, we thank Oleg
  Tsai et al.\ for kindly sharing their test beam slot.

\end{thebibliography}

\end{document}